# Navigating the Ethical and Societal Impacts of Generative AI in Higher Computing Education

Janice Mak*
Arizona State University
Arizona, USA
Janice.Mak@asu.edu

Joyce Nakatumba-Nabende*
Makerere University
Kampala, Uganda
joyce.nabende@mak.ac.ug

Tony Clear*
Auckland University of Technology
Auckland, New Zealand
tony.clear@aut.ac.nz

Alison Clear*
Eastern Institute of Technology
Auckland, New Zealand
AClear@eit.ac.nz

Ibrahim Albluwi
Princess Sumaya University for Technology
Amman, Jordan
i.albluwi@psut.edu.jo

Oana Andrei
University of Glasgow
Glasgow, United Kingdom
oana.andrei@glasgow.ac.uk

Lorenzo Angeli
University of Trento
Trento, Italy
lorenzo.angeli@unitn.it

Stephen MacNeil
Temple University
Pennsylvania, USA
stephen.macneil@temple.edu

Solomon Sunday Oyelere
University of Exeter
Exeter, United Kingdom
s.oyelere@exeter.ac.uk

Matthew Hale Rattigan
University of Massachusetts Amherst
Massachusetts, USA
rattigan@cs.umass.edu

Judy Sheard
Monash University
Melbourne, Australia
judy.sheard@monash.edu

Tingting Zhu
University of Toronto
Mississauga, Canada
tingting.zhu@utoronto.ca

## Abstract

Generative AI (GenAI) presents societal and ethical challenges related to equity, academic integrity, bias, and data provenance. In this paper, we outline the goals, methodology and deliverables of their collaborative research, considering the ethical and societal impacts of GenAI in higher computing education. A systematic literature review that addresses a wide set of issues and topics covering the rapidly emerging technology of GenAI from the perspective of its ethical and societal impacts is presented. This paper then presents an evaluation of a broad international review of a set of university adoption, guidelines, and policies related to the use of GenAI and the implications for computing education. The Ethical and Societal Impacts-Framework (ESI-Framework), derived from the literature and policy review and evaluation, outlines the ethical and societal impacts of GenAI in computing education. This work synthesizes existing research and considers the implications for computing higher education. Educators, computing professionals and policy makers facing dilemmas related to the integration of GenAI in their respective contexts may use this framework to guide decision-making in the age of GenAI.

*Working Group Co-leader. All authors contributed to this research.

## CCS Concepts

• **Social and professional topics → Computer science education**.

## Keywords

Generative AI, Higher Computing Education, Ethical Impacts, Societal Impacts

## 1 Introduction

The need to incorporate an ethical and societal perspective into higher computing education's adoption of Generative AI (GenAI) technologies is pressing. Moshe Vardi (former CACM editor [348] has bewailed the discipline and profession's disturbing absence of concerns over ethics. In adding a higher computing education perspective to these concerns we echo the views of Biesta [42] *"We must expand our views about the interrelations among research, policy, and practice. Furthermore we need to keep in view education, as a thoroughly moral and political practice, requires continuous democratic contestation and deliberation"* [42]. GenAI has a wide range of impacts on how we access and use information, particularly in educational settings [114]. These impacts extend to society and include impacts on intellectual and creative works and the potential infringement of authorship. Differences in institutional GenAI policies (and in funding) may create unequal access to AI tools, the potential disparity in student knowledge of AI tools, responsible uses of AI tools, ethical questions about AI tools, and uneven student knowledge of the benefits and limitations of AI tools. GenAI introduces questions concerning academic integrity, bias, and data provenance [78, 81]. The training data's source, reliability, veracity, and trustworthiness may be in doubt, creating broader societal concerns about the output of the GenAI models.

Given the rapid proliferation and adoption of GenAI, along with its ethical and societal impacts, and the critical role that higher education computing plays at the intersection of education, industry, digital governance, and socio-technical ecosystem, we aim to answer the following research questions:

**RQ1** In what ways does GenAI impact higher computing education through the lens of ethical and societal impact?

**RQ2** In what ways does GenAI affect the socio-technical dynamics of higher computing education institutions?

**RQ3** What are the implications (e.g. challenges, opportunities, limitations) of integrating GenAI in higher computing education?

As an Innovation and Technology in Computer Science Education (ITiCSE) working group, our collaborative research team, comprised of 12 individuals from eight different countries (Australia, Canada, Italy, Jordan, New Zealand, Uganda, the United Kingdom, and the United States), was organized into three collaborative subgroups ("literature review", "policy evaluation", and "framework formulation"; see Fig. 1). The research questions above are explored through the development of the three key deliverables of this paper:

- A **systematic literature review** that addressed a wide set of issues and topics related to GenAI from the perspective of its ethical and societal impacts in the context of computing education. Key databases (ACM, IEEE, Scopus) were initially searched for terms including higher education, computing education, and GenAI. After filtering and focused reading, a final set of papers were then analyzed in depth to extract and report their ethical and societal impacts.
- A global perspective and **evaluation of university policies** on the adoption and guidelines for use of GenAI for computing teaching, education and research.
- The **Ethical and Societal Impacts Framework** (ESI- Framework), a metacognitive guide to analyze, probe, and interrogate the ethical dilemmas associated with the use of GenAI in higher education. We intend for the ESI-Framework to inform practice and guide decision-making for educators, computing professionals, and policy makers as they navigate the integration of GenAI in their contexts.

As a working group, we were supported through the Association for Computing Machinery Education Advisory Committee's (EAC) taskforce on the Ethical and Societal Impacts of GenAI in Higher Education. We align our work on this paper with the goals of this ACM EAC taskforce, which has hosted quarterly webinars related to current issues, questions, and approaches that higher education institutions have taken to address the ethical and societal impacts of computing [352]. We also connect this paper to the the ACM Code of Ethics [119].

## 2 Literature Review

We conducted a systematic literature review (SLR) [188] to address the growing use and integration of GenAI in computing education from the perspective of its ethical and social impacts. Although cognitive and pedagogical aspects have been investigated in previous working groups [278, 280], these ethical and societal impacts have received limited attention in the literature to date [68]. Yet, their breadth is very wide: cf. for example the *computing ethical education* framework developed by Martin et al. [227], and the broad range of topics concerning *computers and society*, classified by Baecker [30] under: *Opportunities: computer applications; Risks: technological threats; Choices: challenges for society.*

Given the breadth of topics and the large number of papers (~2500) on GenAI in computing education, framing the literature review has posed challenges. A guiding protocol has necessarily evolved as the work progressed by consensus between the researchers in the literature review subgroup.

### 2.1 Methods

We posed three research questions, of which RQ1 and RQ3 are addressed in this literature review section.

To execute our literature review, we engaged in the following process: 1) developed a search string, 2) selected databases to search, 3) applied automated screening based on page count, 4) defined a filtering approach based on inclusion and exclusion criteria, 5) developed a coding scheme, and then 6) coded the research papers.

Given the breadth of the literature review and the ambiguity of how ethical issues are discussed, we faced a tension between needing to retrieve a broad and representative sample while ensuring that the filtering process remained tractable. For example, using overly narrow terms, like 'ethics' and 'society,' would potentially miss relevant papers that use more nuanced terms to describe these concepts. However, having a less constrained search term might produce too many papers to review manually. Wohlin [360] further describes this tension between these two competing priorities:

> - Very difficult to formulate good search strings, terminology used is not standardised and if using broad search terms, a large number of irrelevant papers will be found in the search.
> - The latter creates substantial manual work that also is error-prone.

Our team decided to err on the side of over-including papers and doing a more extensive manual filtering of the returned papers. Inspired by a recent working group [280] that conducted a systematic literature review of GenAI, we scoped our search using the following criteria:

- **Domain**: Computing education
- **Topic**: Generative AI
- **Context**: Higher education
- **Focus**: Ethical and societal impacts

We searched for papers based on the first two criteria (i.e., domain and topic) and then filtered those papers based on the last two criteria (i.e., context and focus). This enabled us to retrieve a broad sample of initial papers, then more carefully remove papers that were not about higher education or did not contain ethical or societal impacts. This open-endedness helped to account for the ambiguity in how these topics are discussed within the papers.

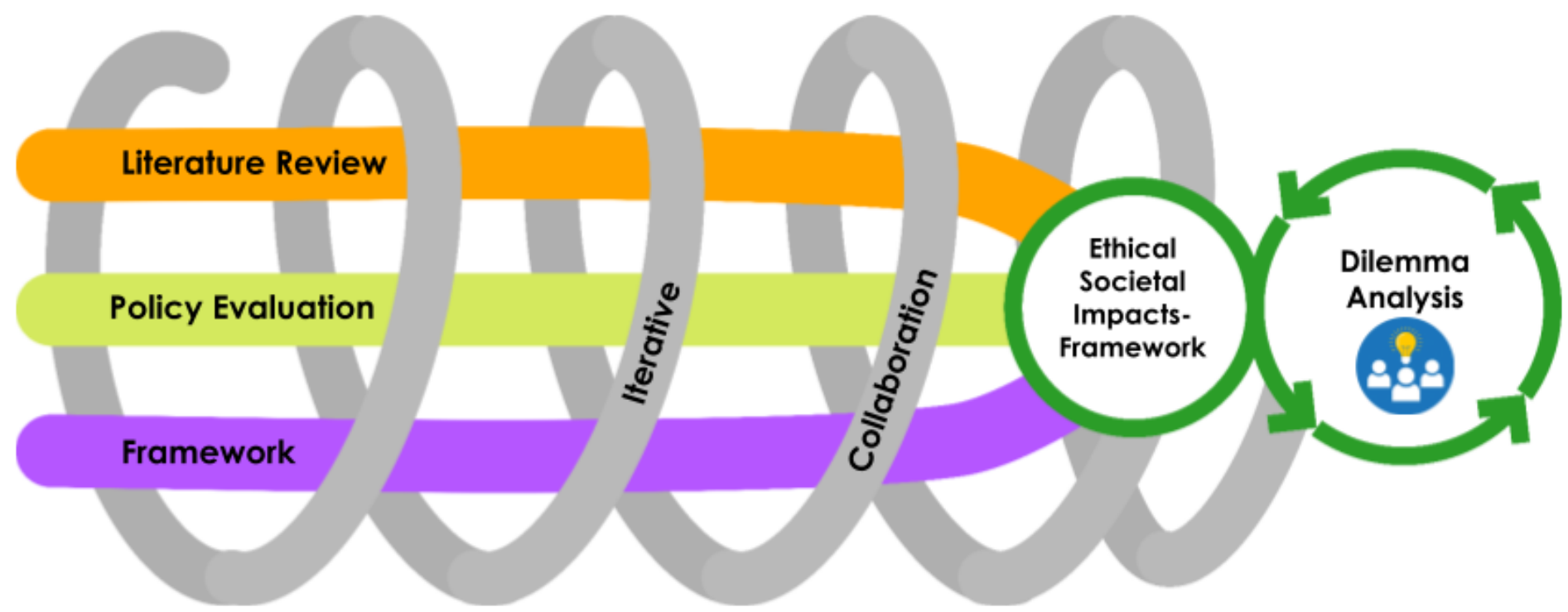

**Figure 1: Working group organization**

*2.1.1 Search String Construction.* We conducted the search string based on the **Domain** and **Topic** to ensure a broad set of initial papers. The final search string focused on the domain and topic using the following keywords for each:

- **Domain:** "computing education" OR "computer science education"
- **Topic:** "generative AI" OR "large language model" OR "llm" OR "chatgpt"
- **Filter on Publication Date:** Since 2020

These were combined to create the following search string:

**Search String** = **Domain** AND **Topic** with FILTER **Publication Date**

This search string partially aligns with the classic PICO approach for systematic literature reviews [36], where the **Domain and Context** effectively addressed **Population** of the PICO elements, and **Topic** addressed the **Intervention and Outcomes** of the PICO elements.

*2.1.2 Search String Validation.* Six key papers [122, 146, 262, 279, 282, 374] were initially identified and served as validation for the search strings. All of the key papers were returned using the above search string.

*2.1.3 Databases.* Aligned with recent systematic literature reviews in computing education research [278, 280, 375], we searched the ACM Digital Library, IEEE Xplore, and Scopus to ensure a broad and representative set of papers. A filter was applied to include papers published in 2020 and after. The search returned a total of 3,829 references from all three databases. The process is outlined in Figure 2.

*2.1.4 Paper Filtering Process.* As shown in Figure 2, the papers were removed if they did not align with the inclusion and exclusion. This filtering process occurred in two stages. In the first stage, papers were marked as 'include' or 'exclude' based only on the abstract and title. At the second stage, papers were removed based on a focused read, where a researcher read the paper to identify whether a code could be applied.

Papers were excluded according to the exclusion criteria (EC):

(1) less than or equal to 3 pages, OR
(2) not written in English, OR
(3) the type of publication is a table of contents or entire conference proceedings, OR
(4) a literature review or policy report, OR
(5) a duplicate between databases.

Papers were included according to the inclusion criteria (IC):

(1) the topic is on the impacts of GenAI, AND
(2) the domain is computing education, AND
(3) the population is higher education.

Papers that were excluded because they were either literature reviews or policy reports, were retained for a separate analysis for developing the framework (Section 4.1).

*2.1.5 Ethical and Social Impact Codes.* An initial set of ethical and societal impact codes were adopted, based on the topics identified by Martin et al. [227]: Quality of life; Use of power; Risks and Reliability; Property Rights; Privacy; Equity and Access; Honesty and Deception.

As the coding process progressed, a further set of inductively-derived ethical and societal codes from our educational context was added to the initial deductive set derived from the earlier work of Martin et al. [227]. These codes emerged after initial code calibration between working group members, and the papers were recoded:

- Computing Ethics and Professionalism
- Sustainability
- Multiple ethical issues
- Maleficence (Negatively impacts learning)
- Beneficence (Positively impacts learning)
- Educational Insights

This evolving definition of the terms relating to ethical and social impacts of GenAI in our context, validated the decision to derive them empirically from the data rather than predetermining the scope through a more tightly defined search string. These new social and ethical codes which emerged reflected: a) the broader educational and professional focus of ethics (rather than single

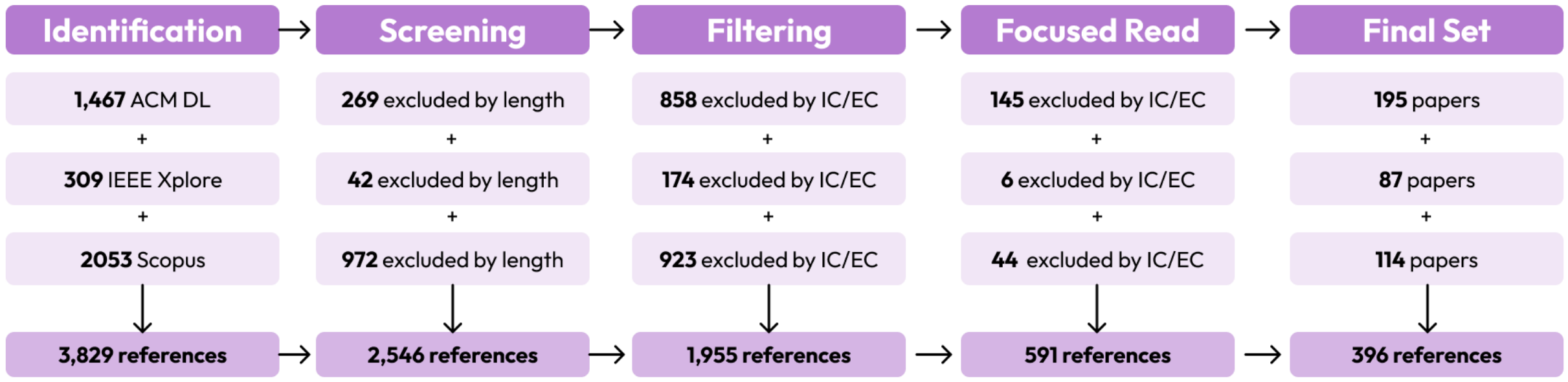

**Figure 2: The paper filtering process during the literature review**

topic) in some papers (such as those focused on ethics courses); b) the increasing concerns about the sustainability of emerging technologies such as GenAI [67]; c) papers which covered multiple ethical and social issues; d) negative impacts on learning; d) positive impacts on learning; and e) insights about how GenAI impacts students and classroom environments. A paper was coded under educational insights if it

- investigates how learners use GenAI tools, OR
- discusses both improvements and impediments to learning, and offers insights into how best to leverage GenAI for education, OR
- has a focus on changes to curriculum, OR
- claims that it saves time for students without other evidence of impact on learning, OR
- discusses implications for students' future careers.

During the coding process, two other themes emerged, including Teaching Resources and Teaching Practices. Nevertheless, these studies did not explicitly discuss the ethical and societal impacts of GenAI, and were thus removed. This resulted in a final set of 293 papers (see Table 11).

*2.1.6 Coding and Extraction.* The full text of papers meeting these criteria were reviewed to extract codes on ethical and societal impacts, courses, and programming languages that were involved in the studies. This set of papers will be referred to as the initial coded set. Only a single code was applied for each paper based on which was deemed most salient. If multiple ethical and societal issues were discussed without a single focus, it was coded 'multiple'. In the case of courses, if a study involved multiple courses, it was coded as 'multiple'. If the course context was not mentioned in the article, it was coded as 'none'.

*2.1.7 Refined Analysis and Extraction.* Since a few literature reviews have focused on the impact of GenAI on teaching and learning in computing education [55, 61, 273, 279, 283, 288, 326], we excluded papers that were coded as 'negatively' or 'positively' impacts learning in the refined analysis and extraction. In addition, we further excluded papers if their discussions on ethics were not based on their empirical data. This set of papers will be referred to as the refined set. We extracted the quotes that directly answer RQ1 and RQ3 from the refined set. Each paper has a primary ethical code, but can be extracted for multiple ethical themes if identified as relevant. For papers coded as *educational insights*, a second iteration of coding was conducted, to extract any further ethical issues explicitly identified in the paper. This analysis is presented under the section on ethical and societal impacts below.

## 2.2 Results

Here we present the results of our analysis of the data collected from the systematic literature review. We first present the analysis of the review studies, then quantitative analysis of the initial coded set, and finally our extraction of ethics from the refined set.

*2.2.1 Other reviews of GenAI in computing education.* During the filtering process of the systematic literature review we identified 94 review studies. These were not included in our SLR final dataset but were considered important to analyze to identify any gaps in the literature. Applying our inclusion criteria:

- **Domain**: Computing education
- **Topic**: Generative AI
- **Context**: Higher education
- **Focus**: Ethical and societal impacts

We filtered these papers based on the first three criteria. This resulted in six review studies. There were a further six reviews where the papers were mostly but not exclusively set in the Higher Education (HE) context. The details of these papers are shown in Table 1.

Of the six reviews that addressed the use of GenAI in computing education in the higher education context, only three discussed ethical issues. This result echoed the concerns about the computing discipline and profession expressed by authors such as Vardi [348] and Ozkaya [260], and more broadly about education as a moral and ethical site by Biesta[42]. The earliest review found was a systematic literature review (SLR) by Prather et al. [278], which investigated the use of GenAI in Higher Education. That SLR resulted in a set of 71 papers from the ACM Digital Library, Taylor & Francis Online, IEEE Xplore databases. The results of the analysis showed a strong emphasis on work reporting the capabilities of the GenAI tools. The authors discuss the potential impact of GenAI on computing education practice and ethical issues raised by the use of GenAI in computing education, framed by the ACM Code of Ethics. An SLR by Cambaz and Zhang [55] resulted in a set of 21 papers from the ACM Digital Library, Scopus, and Google Scholar. The review focused on the teaching and learning practices in programming education that utilized LLM-based code generation models. The review found that the most commonly reported uses of LLM-based code generation models by teachers were generating assignments and evaluating student work, while for students, the models function as virtual tutors. Regarding ethical concerns, the review identified risks to academic integrity from misuse of the models and risks for learning due to errors in generated content or over-reliance on the models by students. The authors conclude that LLM-based code generation models can be an assistive tool for both learners and instructors if the risks are mitigated. An SLR by Liu et al. [208] resulted in 46 papers from Web of Science (WoS), Springer, Google Scholar. The review explored the educational applications and research on AI-assisted programming technology. The findings provide evidence for the value of AI code-generation tools while also showing their technical limitations and ethical risks.

Two reviews of GenAI in computing education in the higher education context discussed societal issues, focusing on the impact of the use of GenAI on learning and the implications for future work in industry. An SLR by Prather et al. [280] resulted in a set of 71 papers from ASEE Peer, arXiv, Scopus, ACM Digital Library, and IEEEXplore. The review investigated the use of GenAI in computing classrooms. The results were triangulated with a survey of educators and industry professionals and interviews with educators and researchers to give an in-depth understanding of what is happening on the ground in computing classrooms. An SLR by Gaitantzi and Kazanidis [108] resulted in 31 empirical studies from Google Scholar and Scopus. The review investigated how AI applications support teaching and learning in computer science, with a focus on database education. The review identified a shift in programming education where changing practices in industry due to the increasing use of AI in software development had prompted "a need to align curricula with evolving industry expectations". The review concluded that "careful integration of AI tools can complement traditional instruction, emphasizing the critical role of human educators in achieving meaningful and effective learning outcomes." However, the review highlighted the limited research in GenAI applications in database-related research.

The remaining review of GenAI in computing higher education was a scoping literature review by Humble [149], resulting in 129 papers from the Web of Science and Scopus databases. A SWOT analysis was used to identify strengths, weaknesses, opportunities, and threats with the use of GenAI for computing education. The findings highlighted several challenges posed by GenAI, such as potential biases, over-reliance, and loss of skills, but also several opportunities, such as increasing motivation, educational transformation, and supporting teaching and learning.

The six other reviews of GenAI in computing education were not set exclusively in the higher education context. None of these discussed ethical or societal issues.

The literature reviews of GenAI in computing education that we found were focused mostly on investigations of programming or software engineering courses. In addition, there has not yet been a literature review that provides an in-depth analysis of the ethical and societal aspects of GenAI. We fill a critical gap by reviewing across different courses and with a focused analysis on ethical and societal impacts.

*2.2.2 Descriptive Statistics.* There is a global disparity in the distribution of scholars contributing to this topic (Figure 3). The US has the highest number of non-unique authors affiliated (550), followed by New Zealand (87), Canada (71), Finland (70), Germany (67), and India (63). Of the 66 countries with at least one author contributing to the field, 41 have fewer than 10 non-unique authors. A significant gap is observed in the Continent of Africa.

Figure 4 summarizes the count of the courses investigated in the initial coded set of 293 studies. The majority of the studies are in Programming or Introductory Programming courses that account for 41% of all the studies. There are 36 papers that did not specify in which courses the studies were carried out, and 22 studies investigated multiple courses. About 7% of the investigations are in Software Engineering and 3% in Data Science courses. Most other courses have a count of 1 to 3.

There are mixed views about the impact of GenAI on computing education, where 71 studies found GenAI mostly contributing to learning, 19 reported negative affects on learning, and 70 captured varied perspectives (Figure 5).

A high percentage of the studies in Introductory Programming, Algorithms and Data Structures, and Computer Science courses reported that the use of GenAI mostly contributes to learning (Figure 6). For studies that reported programming languages, the majority involve Python, followed by Java (Figure 4). Studies in Software Engineering and Programming courses had equal contributions to two major themes, mostly contributing to learning and educational insights. Similarly, investigations that are based on multiple courses also map to these two categories. Studies that did not report specific courses investigated multiple ethical and societal impacts, including honesty and deception, risks and reliability, mostly contributing to learning, educational insights, as well as multiple themes. The theme on 'negatively affecting learning' was reported from Programming and Intro Programming courses.

*2.2.3 Dilemmas in Computing Education.* Synthesizing these studies tallied in figure 5, dilemmas faced by students, educators, and institutions surface, and tensions emerge within individual decisions or between different communities.

Students use GenAI for a variety of tasks, including content generation, problem-solving, debugging, conceptual understanding, and exam preparation [25, 308]. For students, personalized learning and increased productivity were reported as the predominant benefits of GenAI [291, 331]. Students are also aware of problems associated with the use of GenAI, such as hallucination, overconfidence, and over-reliance [308]. Regarding honesty and deception, students may make a conscious decision to commit an academic offense when they are under significant pressure to meet deadline [25]. This leads to a tension between students' decisions and institutional policies, as a study reported that 50% of anonymized students admitted to including AI-written code for assignment submissions even when AI use was prohibited [291].

Educators see both advantages and disadvantages associated with GenAI in computing education. The advantages are opportunities for higher levels of learning outcomes, personalized learning, timely feedback, and alternative assistance [155, 357, 374]. Nevertheless, if students do not use these tools correctly, it will create disadvantages in their further studies or competitiveness in their future workplace. To ensure that students have the skills to succeed in their future journey, educators either came up with solutions to detect the misuse of GenAI, or modify learning outcomes [291].

**Table 1: Review Studies of GenAI in Computing Education**

| Year | Title | HE/HE+ | Method | Papers | Cite |
|---|---|---|---|---|---|
| 2023 | The Robots Are Here: Navigating the Generative AI Revolution in Computing Education | HE | SLR | 71 | [278] |
| 2024 | Exploring Human-Centered Approaches in Generative AI and Introductory Programming Research: A Scoping Review | HE+ | Scoping Review | 28 | [326] |
| 2024 | Navigating the Pitfalls: Analyzing the Behavior of LLMs as a Coding Assistant for Computer Science Students | HE+ | SLR | 72 | [273] |
| 2024 | Risk management strategy for generative AI in computing education: how to handle the strengths, weaknesses, opportunities, and threats? | HE | Scoping Review | 129 | [149] |
| 2024 | Toward Artificial Intelligence-Human Paired Programming: A Review of the Educational Applications and Research on Artificial Intelligence Code-Generation Tools | HE | SLR | 46 | [207] |
| 2024 | Use of AI-driven Code Generation Models in Teaching and Learning Programming: a Systematic Literature Review | HE | SLR | 21 | [55] |
| 2025 | A Bibliometric Exposition and Review on Leveraging LLMs for Programming Education | HE+ | Bibliometric | 195 | [283] |
| 2025 | A Systematic Literature Review of the Opportunities and Advantages for AIGC (OpenAI ChatGPT, Copilot, Codex) in Programming Course | HE+ | SLR | 24 | [61] |
| 2025 | Beyond the Hype: A Comprehensive Review of Current Trends in Generative AI Research, Teaching Practices, and Tools | HE | SLR | 71 | [280] |
| 2025 | Generative AI in Computer Science Education: Insights from Topic Modeling and Text Network Analysis | HE+ | Literature Review | 151 | [10] |
| 2025 | Large Language Models in Computer Science Education: A Systematic Literature Review | HE+ | SLR | 125 | [288] |
| 2025 | The Role of Artificial Intelligence in Computer Science Education: A Systematic Review with a Focus on Database Instruction | HE | SLR | 31 | [108] |

[1] HE indicates Higher Education and HE+ indicates the review was not exclusively in Higher Education

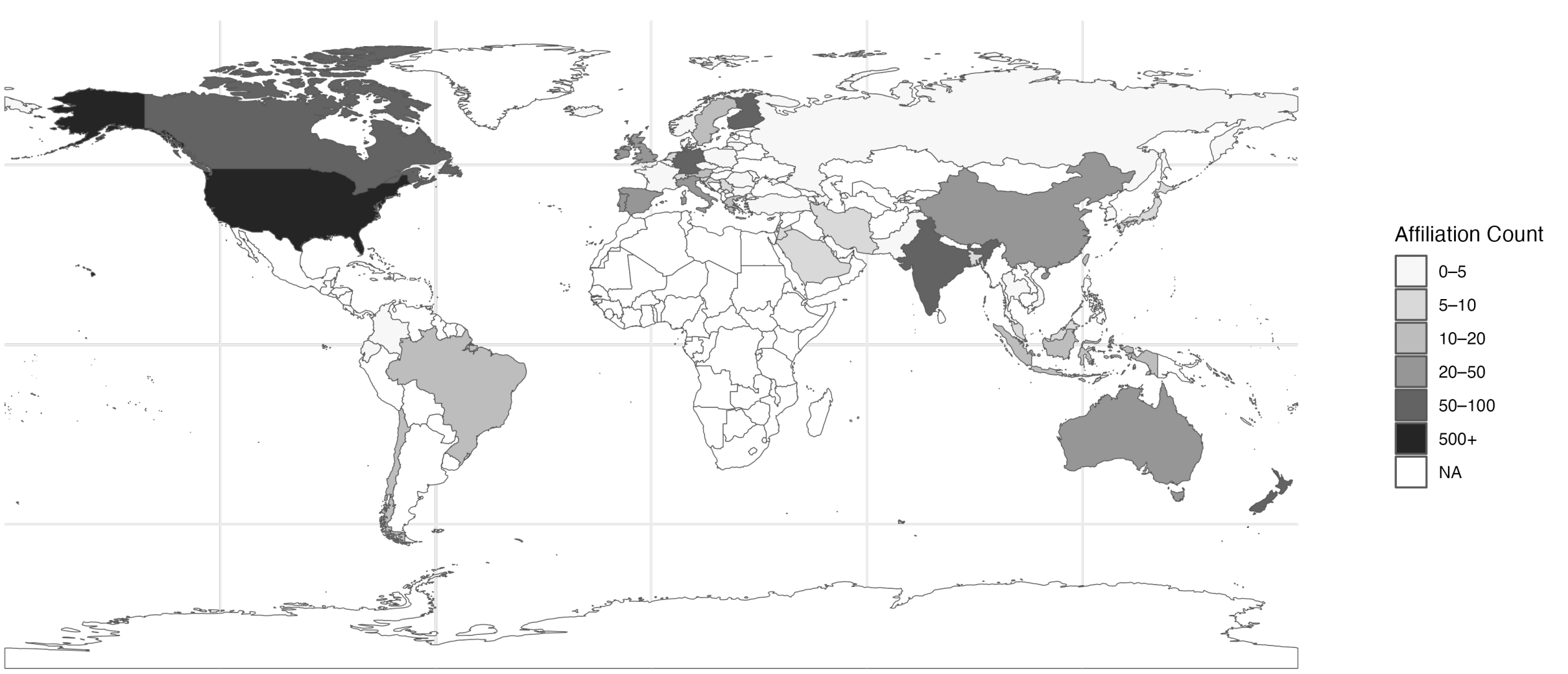

**Figure 3: The geographical distribution of author affiliations.**

However, the current AI detection tools are not 100% reliable. If students are wrongly accused, other ethical and societal issues arise. However, if students' inappropriate behaviors are not corrected and they do not achieve the learning outcomes (which is now harder to evaluate), they will face consequences later down the road. Currently, many educators do not teach students how to correctly and responsibly use GenAI, or they lack the tools or did not find time to teach students these aspects [291].

While GenAI provides multiple opportunities for teaching and learning, dilemmas and tensions arise. It cuts across different social levels Martin et al. [227], radiating from the individuals, to student groups, classrooms, institutions, professional workplaces, and a broader society. Implications can be different for the current status compared to the future status. A framework needs to be developed to guide individuals or institutions to make informed and responsible decisions, and consider how current actions may lead to future consequences.

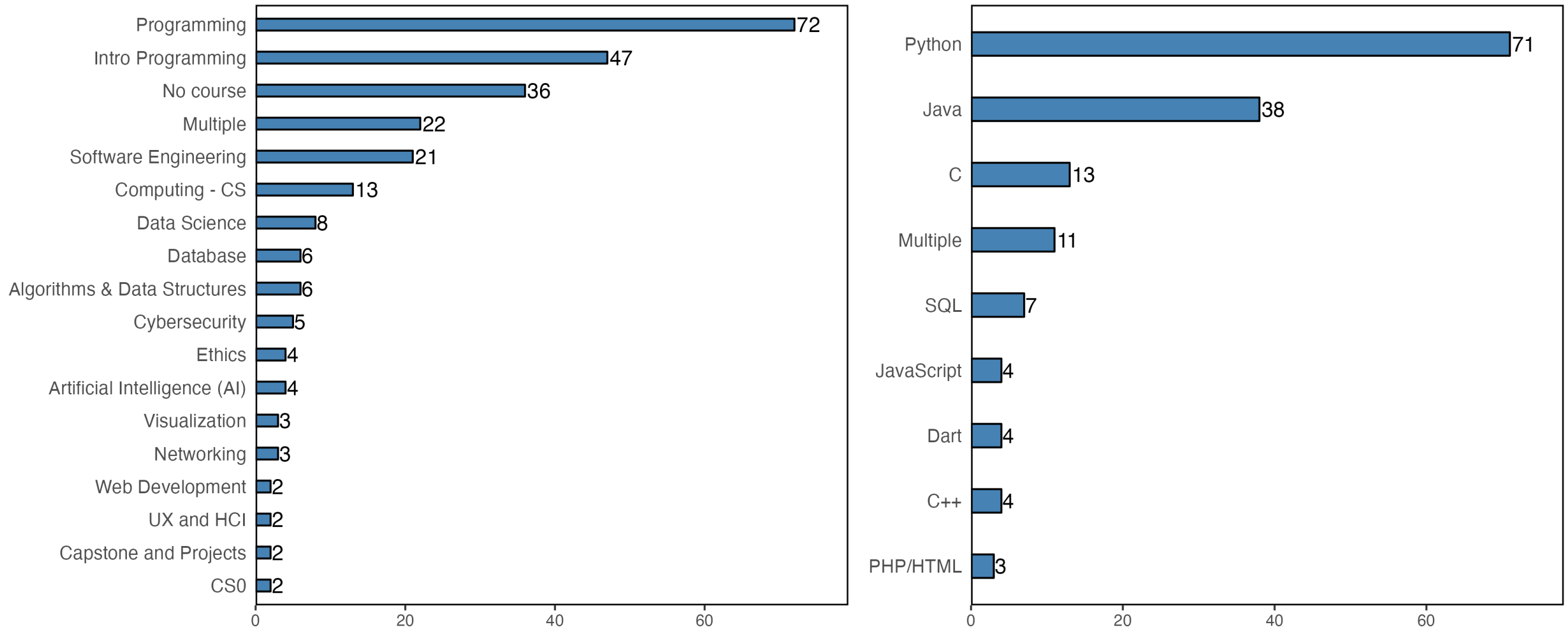

**Figure 4: The bar chart on the left shows the frequency counts of courses. The bar chart on the right shows the frequency counts of programming languages. Only those with frequency count larger than 1 are included.**

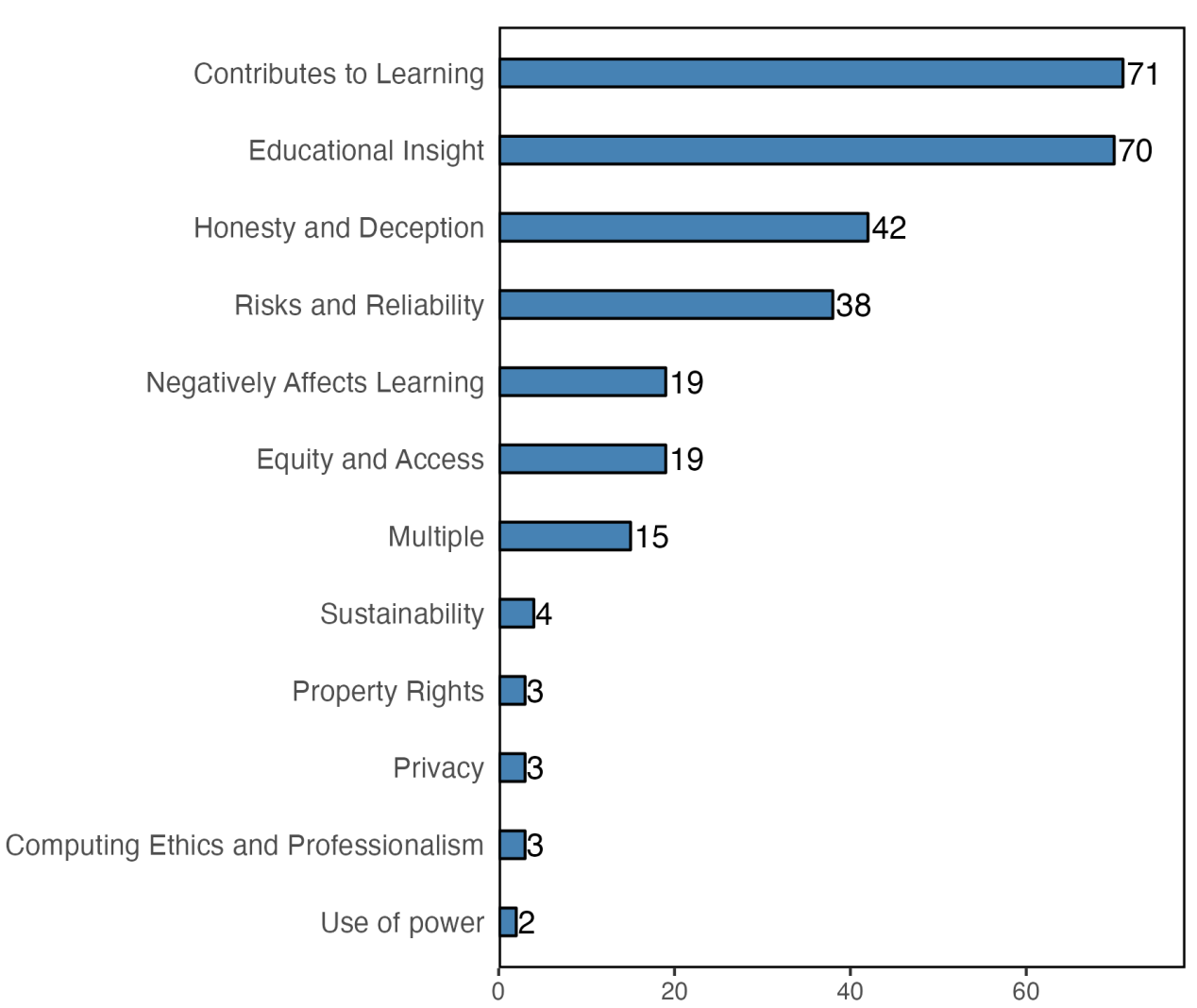

**Figure 5: A barchart showing the frequency counts of each code. Only one code was applied per paper.**

2.2.4 *What Are the Ethical and Societal Impacts?* Most of the analyzed papers in the refined set focused on the impacts (elaborated below) on honesty and deception, while fewer papers contributed to the understanding of how GenAI impacted equity and access, risks and reliability, sustainability, computing ethics and professionalism, use of power, privacy, and property rights (Figure 5). This section presents results from the in-depth coding of the refined set of societal and ethical related papers (Table 2).

| Ethical Concern | Citation |
|---|---|
| Honesty and Deception | [32, 40, 41, 43, 49, 51, 62, 83, 98, 99, 104, 126, 129, 130, 143, 161, 164, 174, 180, 184, 198, 202, 221, 224, 256, 264, 265, 267, 269, 279, 289, 291, 294, 296, 299, 303–305, 316, 324, 332, 354, 355, 368, 374] |
| Risks and Reliability | [16, 35, 80, 178, 206, 218, 282] |
| Equity and Access | [39, 146, 165, 279, 281, 349, 374] |
| Privacy | [3, 54, 276, 308, 324, 347] |
| Property Rights | [54, 194, 313, 324] |
| Use of Power | [181, 346] |
| Sustainability | [242, 244] |
| Computing Ethics and Professionalism | [54, 93, 293] |

**Table 2: List of papers included for refined ethics analysis.**

**Honesty and Deception.** The largest number of papers (45) covered aspects of honesty or deception, and most did so in the context of academic integrity and cheating. Many papers discussed student use of GenAI for learning and assessment, with unauthorized use of GenAI tools and increasing academic integrity incidences impacting learning and fairness. From the teaching perspective, a number of papers discussed the challenges that GenAI presents to teaching programs, particularly regarding how to incorporate GenAI into teaching and assessment and how to discourage the increased levels of cheating that the GenAI tools afford. A number of papers investigated the potential of AI for making cheating easier and the challenges in detecting cheating.

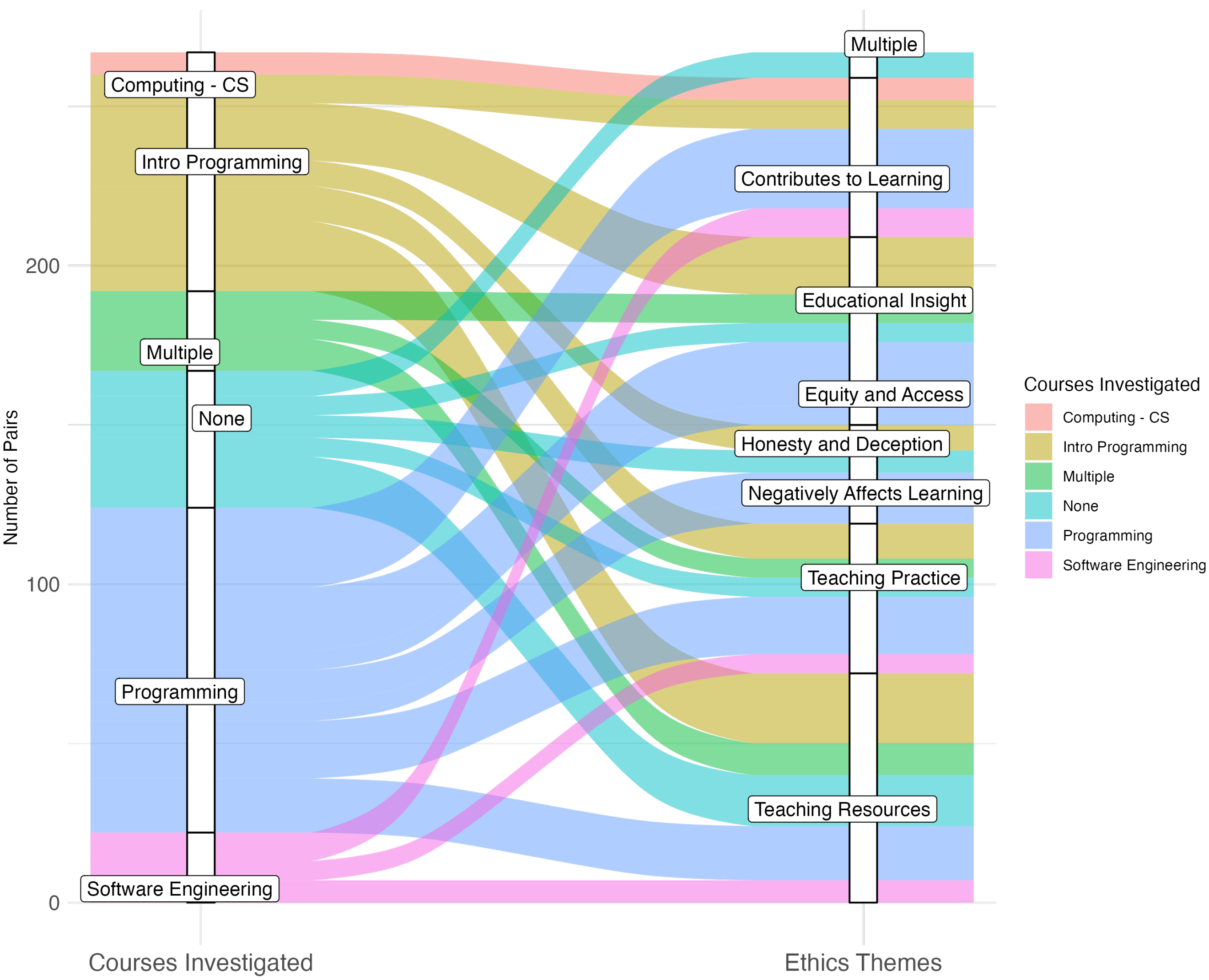

**Figure 6: A Sankey diagram showing the mapping between courses and ethics. The edges represent connections with more than 3 pairs.**

While acknowledging benefits to students' learning from productive uses of GenAI, unauthorized use of the tools for learning tasks and assessment was of widespread concern [278]. There was a general perception and evidence from a number of studies that, since the advent of GenAI, there has been a rise in incidences of academic dishonesty. With the free and accessible GenAI tools, students now have an easy way to produce solutions that cannot be detected using common approaches [41]. Richards et al. [296] compared this situation with the use of essay mills, proposing that:

> ...the deployment of ChatGPT provides a new and distinct method for cheating, drives the financial cost to zero, greatly reduces the likelihood of being accused of plagiarizing content, and has reduced the time needed to produce scripts to a few minutes. Unlike essay mills, the speed and availability of ChatGPT allows its use in remote examination environments. (p.5:27)

GenAI presents students with ethical dilemmas about how and when to use the tools for their learning and assessment. Several papers illustrate these tensions. A study involving interviews of instructors and students by Zastudil et al. [374] found that all instructors expressed concerns about students using GenAI tools inappropriately, while students reported their willingness to use GenAI to cheat on an assignment when they don't see the value of the assignment. In another study by Randall et al. [291], a survey of software engineering students found that 47% of the surveyed students claimed they would likely use GenAI tools for university work even if their usage had been banned.

A lack of clarity about the acceptability of practices further exacerbates these dilemmas. A survey of computer science students

(n=70) by Rogers et al. [299] found that a concerning number of students believe it is acceptable to use ChatGPT to provide answers in exams (20%) or solutions for assignments (41%) – practices that are not permissible in many computing education contexts. Surveys by Keuning et al. [184] over three consecutive teaching periods found that an increasing number of students considered it ethical to auto-generate and submit an entire assignment without understanding it. Whereas a study (n=57) by Prather et al. [278] found consensus among academics that submitting an auto-generated solution without understanding to be unethical; however, there were differing opinions about the ethics of this practice when students possessed a full understanding of the generated code or wrote the code in a different language and then translated it into the language of instruction.

Students are further challenged by the nature of the GenAI tools, which do not provide information on the source of the information that is retrieved. As a result, students who use the tools for their assessable work are at risk of accidental plagiarism [304]. A study by Bikanga Ada [43] found that over half the respondents were concerned about accidentally plagiarizing when using ChatGPT.

Inappropriate or dishonest use of GenAI poses considerable risks to student learning. While tools such as ChatGPT and Codex offer instant solutions, by using these tools to solve learning tasks students avoid essential learning experiences. When producing AI-generated answers without engaging in analysis or reflection, students miss opportunities to develop their skills and knowledge, and to receive meaningful feedback on their progress [40, 224]. This practice creates disparities in skill development and undermines the authenticity of assessments, as performance may not reflect genuine understanding [332]. A study by Chen et al. [62] found statistically significant learning losses with students using GenAI in assessments, which was proportional to the amount of presumed plagiarism.

Of further concern is that excessive reliance on GenAI can lead to a cycle of self-deception. Students who use GenAI for learning and assessment tasks may perceive themselves as capable while actually failing to acquire foundational knowledge and skills. Furthermore, the readily available solutions may mean that students are not motivated to search for alternative solutions, thus limiting their learning to what is offered by the GenAI tool, which, in fact, could be inaccurate or misleading. This not only impacts their learning but has implications for their future professional work, particularly in areas such as programming where an ability for independent reasoning is critical [278]. The ease with which the GenAI tools provide solutions further exacerbates these problems. [130].

Cognitive science research has relevance in supporting findings here, through work observing the risks of technological assistance causing skill decay in complex tasks demanding controlled processes [216]. Macnamara et al. [216] have observed from work with airline pilots that AI assistants may cause greater decrease in skill than traditional automation systems, as the AI assistants *are designed to mimic cognitive skills—that is, they recognize patterns, reason about potential outcomes, and often guide the user to a specific action*, therefore regular engagement with an AI assistant reduces the opportunity to keep skills honed. Furthermore, learners who have engaged with AI assistants in their skill development may believe they have a greater understanding of the task than they do possess. So the over reliance on AI assistants poses risks of self-deception.

Yet there were indications from a couple of studies that some students were aware of the dangers of over-dependence on GenAI tools. A survey of students (n=80 ) by Palacios-Alonso et al. [265] found that students felt guilty or concerned about using GenAI tools, acknowledging that they were cheating themselves and recognized that excessive use of these tools could jeopardize their future employment. A study by Bikanga Ada [43] found that students were aware of these ethical dilemmas posed by GenAI tools and wanted more comprehensive policies and educational programs to help navigate these challenges.

Dishonest use of GenAI tools is potentially harmful to both students and institutions. Students who use GenAI dishonestly to generate solutions for assessment tasks or to assist with exams may achieve better results than their peers who complete tasks independently [174]. Such practices undermine the intended learning objectives of assessments, as students may pass a course by mastering the use of GenAI tools rather than through understanding the content [43]. This creates an imbalance between effort and outcomes, creating disparities in academic results, and disadvantaging students who commit time and effort to genuine learning. Honest students, therefore, face unfair competition, while dishonest students limit their own skill development [98]. These outcomes compromise fairness, bias academic performance, and erode the integrity of assessment systems, bringing potential harm not only to individuals but also to institutions and the wider community [316].

With the widespread acceptance and use of GenAI in industry, the integration of GenAI tools into computing curricula is now viewed by many as essential. In a survey of computer science majors by Smith et al. [324], students perceived that mastering GenAI tools would be vital for their future career competitiveness. However, incorporating these tools into a teaching program requires careful attention to risks to academic integrity and to the impact on learning through dishonest use of the tools. In a study by Rajabi et al. [289] that explored the views of students and faculty members regarding ChatGPT's use, the absence of guidelines from the university and instructors on what would determine misuse of ChatGPT was a recurring theme.

Several papers described approaches that educators have made to safeguard academic integrity, such as weighting invigilated assessment more heavily, banning GenAI use in specific contexts, and explicitly demonstrating to students both the capabilities and limitations of these tools [40, 198]. Others have shifted the focus of assessment from product to process and have included interviews or presentations to determine students' depth of understanding [180, 304, 332].

Embedding principles of transparency in the use of GenAI is critical to maintaining academic integrity. Requiring that students declare when the production of their work has been assisted by GenAI and cite generated outputs appropriately helps preserve fairness while fostering ethical and responsible use [32, 180, 264]. By foregrounding academic integrity in AI integration, educators can prepare students to engage with these technologies ethically, ensuring that GenAI supports, rather than undermines learning and assessment.

Detecting when GenAI tools had been used inappropriately or dishonestly to assist in the development of a solution for a learning or assessment task was a dominant theme, with more than a third of the papers reporting investigations of the performance of GenAI tools in the context of academic integrity. A number of studies investigated the effectiveness of AI tools in producing solutions that could bypass detection tools (e.g., [49, 104, 256, 267, 303, 354]). Other studies investigated methods and tools to detect when GenAI tools had been used to produce content [224, 332]. These studies highlighted both the difficulty of designing assessments that could not be readily solved by GenAI tools and the difficulty in detecting GenAI produced solutions. Tang et al. [332] proposes that the:

> ...impressive proficiency of ChatGPT in generating solutions for intricate exercises highlights the difficulty in crafting exercises meant to challenge AI generated solutions. Nevertheless, the excessively "perfect" nature of AI-generated solutions, particularly those utilizing advanced practices and features, may serve as a distinguishing factor from those of average elementary programmers. (p.482)

Of further (and deeper) concern is that focusing on detection tools may divert attention away from conversations with students about academic integrity that can provide more meaningful support to their learning and development and helping them understand the ethical implications of the use of the GenAI tools.

In support of taking an educative rather than a technological approach to the dishonest practices around the use of GenAI tools, Kendon et al. [180] argues:

> Relying solely on such [detection] tools inserts educators into an arms race with students, in which neither side can trust the other, and the winner is simply the one who brings the greatest and most advanced tools to bear.

In conclusion, GenAI presents profound ethical challenges to computing students and educators. Establishing guidelines and policies for its use in computing programs is essential to ensure the tools are used in a way that benefits learning and maintains academic integrity.

**Use of Power.** Only two papers were tagged as Use of Power. In both cases, power was not explicitly stated, but focused on issues of power. In the first paper, Vahid et al. [346] investigated methods for automatically detecting unsanctioned use of GenAI. While these tools can flag potential academic dishonesty, instructors must still verify cases through time-consuming interviews. Although such reviews act as safeguards, they also shift initial authority to detection algorithms, raising concerns that students may be wrongly accused. In the second paper, Kerney [181] presents tactics such as 'laying traps' to catch when students use AI in ways that go against the course policies. It also presented students counter-tactics to subvert detection by 'Stealthing' their use of GenAI, which in some cases could fool AI detection algorithms. Together, these two papers show how both students and instructors are adapting their practices in response to shifting power dynamics introduced by AI.

**Access and Equity.** Some articles discussed barriers to access or equity issues as they relate to GenAI. For example, Zastudil et al. [374] conducted an early interview study of perceptions of students and instructors, making connections to the digital divide that occurred in the early 1990s when the internet became accessible to some, but not to all. While they merely hypothesized this possibility at the time, multiple studies showed evidence of bimodal distributions of GenAI usage [146, 279], with some students using GenAI tools every day and many others never having used these tools.

Another topic that emerged related to access is the ability for GenAI to support students in their native languages [165, 281]. Jordan et al. [165] prompted a language model to produce introductory programming problems in English, Tamil, Spanish, and Vietnamese. They found that minor corrections were often needed before giving assignments to students, but it shows the potential for GenAI to be used to support students in languages besides English.

There were also papers that explored the use of GenAI to produce more culturally or personally relevant learning materials for students [39, 349]. These included a study by Bernstein et al. [39] where they investigated the analogies that students created to explain computing concepts. A second study by Villegas Molina et al. [349] investigated how large language models could be used to produce a culturally-relevant CS1 textbook for Latines.

Overall, the findings highlight both the challenge of growing disparities in GenAI usage among students and the opportunity presented by emerging efforts to design more culturally and personally relevant computing education. Programming is now also becoming more accessible to students who speak English as a foreign language. Any student may now work on assignments and view explanations in their native languages.

**Risks and reliability**. Most instances that addressed risks and reliabilities focused on the potential for GenAI to produce incorrect information. Given the prevalence of these hallucinations, some papers investigated students' trust in GenAI [16]. Other risks included overreliance, with students potentially using AI to avoid the hard work of learning [35], or negative impacts on students metacognition. Prather et al. [282] showed through an eye tracking study that using AI can help high performing students, but can also negatively impact their metacognitive plans; in some cases leading them to solve the wrong problem [282]. Their work also observed an illusion of competence in which students believed they were more capable of solving problems than they really were. As observed by Macnamara et al. [216] from cognitive science, these *illusions of understanding* are inherent risks of AI assistants. In conclusion, these studies show that GenAI needs to have some forms of guardrails to ensure that students stay on track and don't misuse AI tools. More recent work beyond this review, has begun investigating the use of guardrails [80, 178, 206] as well as tools to scaffold the use of copilot assistants [218]. These provide a promising first step toward addressing some of the risks and reliabilities students face when using AI tools.

**Privacy.** CS students are concerned about privacy when using GenAI tools [308, 324]. In an introductory programming course, students expressed concerns about privacy when using Github Copilot [278]. To mitigate this concern, educators have made some efforts. In a first-year undergraduate relational database course, Pozdniakov et al. [276] evaluated a tool named Feedback Copilot to provide quality feedback and improve students' performance. To protect students' privacy, students' names were replaced by random identifiers before using GenAI models. By aligning with the privacy standards, Pozdniakov et al. [276] found that it provides a user centric application that offers pedagogically sound and ethically responsible approaches. Another recommendation was to "instruct students not to share any personal or confidential details with the LLM during their interactions" [3]. According to a syllabus analysis, some ethics topics, including privacy, have been introduced in CS education [54]. Nevertheless, when attempting to address different dimensions of ethics, the problem becomes complicated. For example, in optimizing sustainability by informing people through new ways of handling data, the privacy is compromised [347].

**Sustainability.** Only four papers investigated or discussed sustainability. Building on their earlier work that interviewed professionals' views on sustainability in Software Engineering, Moreira et al. [244] identified challenges and effective ways to integrate sustainability into Software Engineering curriculum. Moore and McCullagh [242] presented how Sustainable and Ethical Computing was designed into the computing curriculum at their university. One learning objective linked to sustainability, and the environmental challenges from the excessive amounts of electricity and water required to support such large models, is to minimize the size of a machine learning model while maintaining its effectiveness [242]. In this process, students are engaged in an ethical debate regarding the trade-off between model size and accuracy.

**Property Rights.** Both students and educators have expressed concerns about the violation of property rights [194, 324]. Smith et al. [324] surveyed all CS majors at their institution and received 116 responses from undergraduate students and 17 responses from graduate students. Out of them, 14 students voiced societal concerns that included intellectual property violations [324]. In a project-based course that integrates GenAI tools, a student expressed that "from an ethical standpoint as an artist, particularly in the way it sources its backlog of imagery", about how "AI image/art generation in general holds a gray area for art theft". [313] In the process of generating CS1 and CS2 quizzes using GenAI, Kumar et al. [194] suggested to consider the copyright and ownership of content and verify whether the material falls under "fair use" exemptions, but they also stated that "We know that generated content relies heavily on the training cases, and we assume that the LLM providers have resolved the copyright issues regarding the training and the generated content." To prepare faculty for the responsible use of LLM, three CS faculty members at a small liberal arts university created a framework for creating a workshop for faculty across disciplines [46]. They advise to "check the source of the information and cite it properly" and "explicitly acknowledge the assistance of LLMs in the creation of your work (sections or parts that included ideas/issues initially identified via LLMs, or tasks achieved such as editing and paraphrasing, or calculations)."

**Computing Ethics and Professionalism.** The overarching code *Computing Ethics and Professionalism* was added to complement the set of more discrete topic-focused codes derived from Martin et al. [227]'s framework). Although only three papers were coded as *Computing Ethics and Professionalism*, they reflected the growing concerns about the ethical implications of AI and GenAI models and the need for educational responses through courses or curricula. The first paper reviewing computing curricula in 17 Brazilian Universities active in teaching Artificial Intelligence [54] argued for the development of critical ethical reasoning as "extremely relevant to the education of future Artificial Intelligence developers", but "concluded that universities are not fulfilling their role adequately". The neglect of ethics as a topic in favor of technical topics and its placement as a voluntary option in many curricula (some 30 per cent of ethics related disciplines being obligatory and the remaining 70 per cent of topics left to students' individual choice), supported the authors' argument about Universities not fulfilling their role.

In the second paper reviewing the CS2023 curriculum. the authors observed an increased emphasis on practical applications of GenAI and the social and ethical implications. But with GenAI as a technology and its enormous potential for impact, came a demand for GenAI practitioners to attend to its ethical and societal aspects, through students studying "real-world AI applications and their implications" [93].

In the third paper reviewing the challenges posed by *Practicing Responsible Innovation in high performance computing (HPC)*, the authors argued for the critical role of ethics in professional behavior, believing "it is far better to have a voice in shaping innovation than to avoid accountability" [293]. The recurring theme of divorce of ethics from the discipline and its neglect is evident, with ethics still being treated and taught as a discipline entirely separate from computer science. They argue that this mental division "can lead computer science students — potential designers of future HPC and AI systems — to neglect the significance of ethics in their field". So, in one encouraging response to these challenges, the CS2023 curriculum has taken the holistic position that, "Societal and ethical issues are intertwined with applications. Rather than separate them out into their own knowledge unit, which could encourage their isolation into separate, dedicated class sessions, these AI curricular guidelines insist that students learn about societal and ethical issues in the context of AI applications" [93].

**Educational Insights** As noted earlier, papers coded as revealing educational insights were further coded to identify explicit mention of ethical and societal concerns. Of the 68 papers coded as educational insights, 39 papers were excluded since they did not have a focus on ethical issues. These excluded papers typically addressed topics such as automating and supporting aspects of teaching and learning and assessment through GenAI based tools and models. Papers reported initiatives that focused on improving the learning of programming or other computing courses implemented through system designs, prototypes and pilot studies. Of the remaining 29 papers the balance of topics was as tabulated below in Table 3.

As previously reported under the coding and extraction section, and as applied more generally when coding ethical and societal topics, only a single code was applied for each paper based on which was deemed most salient. This tended to understate the prevalence of some codes, with more than one code appearing in several cases, the most commonly occurring pairings involved *risks and reliability*, *equity and access*, *honesty and deception*, and *property rights*, which alternated in salience. Selecting the primary code seemed preferable to simply coding the paper as *multiple ethical issues*.

The one such instance, however, reported a varied and highly critical set of student comments: *"6 of the 13 categories only refer to negative aspects students experience using ChatGPT. Among them*

| Educational Insights - Ethical Concerns | Count |
|---:|:---:|
| Quality of life | 1 |
| Use of power | 3 |
| Risks and Reliability | 9 |
| Property Rights | 1 |
| Privacy | 1 |
| Equity and Access | 3 |
| Honesty and Deception | 8 |
| Computing Ethics and Professionalism | 2 |
| Sustainability | 0 |
| Multiple ethical issues | 1 |

**Table 3: Educational Insights – Ethical Concerns Table**

*are privacy concerns, ChatGPT's overconfidence despite producing incorrect results, hallucinations, and a lack of integrity (e.g., by citing non-existent sources). Moreover, students expressed the necessity to remain critical of the GenAI tool, and to always verify the generated responses. Some students (26) even mentioned the risk of depending too heavily on ChatGPT, causing them to not thoroughly attempt to solve programming exercises themselves anymore"* [308].

In one interesting observation in the context of *quality of life* and GenAI assisted pair programming, students *"missed the dynamic interaction, empathy, and engagement of traditional student-student pairing"* [322].

In the process of coding Educational Insights papers against the ethical topic of *Risks and Reliability* it became clear that in addition to the reliability issues of quality of output, errors and hallucinations, the code covered an underlying ethical risk relating to the use of GenAI in education. This evolving ethical code, which poses a risk to learning, could be termed *Ethical Learning* and can be explained in a student comment: *Worries About Developing Dependence and Hindering Learning: Several participants expressed worries about becoming overly dependent on ChatGPT... P1 cautioned about the "over-reliance on AI", highlighting a tendency to "let it do it for you"...implying a trade-off between efficiency and educational depth. Participants were concerned such dependency potentially harms their learning*[187]. While time precludes fully unpacking this evolving code here, it suggests an important area for further investigation.

## 3 Analysis of University Policies

This section, along with Section 4, addresses the second research question:

**RQ2** In what ways does GenAI affect the socio-technical dynamics of higher computing education institutions?

We examine institutional policies as socio-technical artifacts that both reflect and shape how universities respond to the proliferation of GenAI. We begin by reviewing previous studies that analyzed university policies on GenAI. We then systematically analyze the policies of 21 universities around the world, highlighting overarching themes and variations in how these universities view GenAI, regulate its use, and educate their communities on its challenges and implications. Finally (in Section 4), we derive a framework that provides structured guidance to educators and policy makers on how to reason about ethical and societal issues when preparing policy documents related to GenAI use in their institutions.

### 3.1 Previous Work

Policy documents, that is, documents that discuss how each university equips itself to address the changing educational landscape with the rise of GenAI, are rich sources of insight into how universities are responding to the rapid emergence of GenAI and navigating its challenges and opportunities. Policy documents not only reflect institutional values and priorities, but also play a role in shaping the experiences of students and faculty members. Therefore, many research studies began analyzing university policy documents as early as 2023, soon after OpenAI's ChatGPT gained widespread use. These studies approached the analysis from various perspectives and focused on different geographical areas.

Several studies focused on top-tier universities according to university ranking systems such as QS World University Rankings[2] [213, 344, 363], Times Higher Education (THE) World University Rankings[3] [15, 112, 243] , Academic Ranking of World Universities (Shanghai Ranking) [4] [73], and U.S. News Best Global Universities Rankings[5] [15]. The rationale for this focus is twofold. First, top-ranked universities often have a strong online presence, which makes accessing their policy documents easier than at other institutions. Second, these institutions are frequently among the first to respond to new challenges, and their actions can influence the practices of smaller or less prominent universities. On the other hand, some studies focused on analyzing university policies in specific regions like Asia [75, 365] and Europe [327], while others focused on specific countries like Denmark [90], South Africa [60], Hong Kong [64], and the UK [24, 358], with the greater share being studies that focused on R1 universities in the USA [12, 17, 142, 229, 353]. Several of these studies published the datasets used in their analysis publicly on the internet [112, 142, 163, 353].

The studies also varied in how they analyzed the policies. Jin et al. [163] analyzed the policies using Diffusion of Innovation Theory [298], which focuses on how innovation is communicated and adopted through cultures and institutions. On the other hand, Driessens and Pischetola [90] and Luo [213] adopted Bacchi's "What's the problem represented to be?" framework [29], which postulates that policies are created based on what policymakers believe is "problematic" and requires change. Hence, the analysis looks at the problems represented in the policies, the presuppositions and assumptions underlying these problem representations, how the problematization is communicated, and what is left "unproblematic". Driessens and Pischetola [90] identified three key problematisations in the analyzed policies: assessment integrity, legality of data, and veracity. Driessens and Pischetola [90] also found that the huge planetary costs of GenAI and the exploitative business models and practices are mostly left unproblematized in the policies. On the other hand, Luo [213] found the main problem represented in the policies to be that students may not submit original work for assessment, where GenAI is viewed as "a type of external assistance separate from the student's independent efforts and intellectual contribution". This problem representation does not acknowledge how GenAI complicates the process of producing original work, and is reflected in the policies as a silence towards the evolving meaning of "originality" for work mediated by technology.

[2]https://www.topuniversities.com/
[3]https://www.timeshighereducation.com/world-university-rankings
[4]https://www.shanghairanking.com/
[5]https://www.usnews.com/education/best-global-universities

Other frameworks used in the analysis include Ubuntu Ethics [60] and Kotter's Change Model [112]. Wilson [358] also discussed the policies using a wide range of theories like: Institutional Theory, Advocacy Coalition Theory, Policy Diffusion Theory, and Punctuated Equilibrium Theory. A notable example is the work by Dabis and Csáki [73], which used Directed Content Analysis to identify four clusters of positive values found in government-issued policy documents on AI ethics: (1) Accountability and responsibility, (2) human agency and oversight, (3) transparency and explainability, and (4) inclusiveness and diversity. These clusters of values were then used to analyze university policies in terms of how their "first response" to GenAI was, and then used the analysis to derive best practices regarding GenAI policies at universities.

While the studies discussed above conducted analyses grounded in specific theoretical frameworks, a larger portion adopted a more flexible and descriptive approach. Many of these studies explored general questions such as: "What do the policies cover?", "What themes are present in the policies?", and "What guidance do the policies provide?" [12, 15, 17, 60, 64, 75, 142, 243, 339, 344, 353, 363]. They also examined contextual aspects, including who issued the policies [327, 344, 365], the timing of their release [60, 64, 344, 365], the intended audience or stakeholders [17, 60, 75, 75, 327], and the referenced GenAI tools [12, 60, 75, 75, 344]. Several studies also analyzed the stance of the institutions towards GenAI (are GenAI tools encouraged, allowed, discouraged, or disallowed) [12, 17, 75, 153, 229, 353]. For example, Wang et al. [353] found that none of the top-100 universities in the US banned the use of GenAI tools and that 57 of these universities defer the decision to the instructor. However, out of these 57 universities, 27 adopted a "stance of Prohibition by Default", where the use of GenAI tools is allowed *only if* the instructor explicitly permits it, otherwise their use is considered a form of academic dishonesty. On the other hand, only a single institution adopted a "stance of Permissibility by Default".

While the "acceptance" (or lack thereof) of GenAI tools can be explicit in the policy, some of the studies looked at the perception [353], narrative [75], sentiment [17, 229], or discourse [12] around GenAI that can be *interpreted* from the policy. For example, Tong et al. [339] conducted automated linguistic analysis on 92 policies and found that the use of second-person pronouns was most common, and that only first-person and collective pronouns correlated with a warmer tone and a higher likelihood of discussing GenAI limitations. Ali et al. [12] also found that 38 of the analyzed course syllabi (i.e., 39%) attributed living characteristics to GenAI (as assistants), with some (N=5) even *personifying* GenAI tools by referring to them as peers or tutors, and using terms like "collaborating with AI" when referring to using GenAI tools.

## 3.2 Conceptual Framework

We draw on Taylor et al.'s [140] conceptual framework for policy analysis that studies policy documents from the perspectives of context, text, and consequences. The policy context refers to the issues, social context, and impetus that led to the need for policy to begin with, accounting for the background and socio-political tensions and environment [37]. Analyzing the policy text refers to the process of critically interrogating the text for its explicit and implicit purposes and values, as well as for how the same policy could be interpreted from the perspective of diverse stakeholders. Policy consequences ask that we consider the impacts of policy implementation, including intended and unintended consequences and challenges [11] the policy presents for a diverse range of stakeholders. How is policy implementation impacted by the ways in which policy users interpret it [302]? Is the policy implemented consistently and effectively? Are there policy implementation challenges that need to be addressed related to people, processes or structure [148]? By analyzing policy documents through the lens of context, text, and consequences, we center both the intent of a given policy as well as the foundational values and interrogate its implementation.

## 3.3 Methods

While, as we have seen in the previous section, there is substantial literature analysing university policies on the use of GenAI, there is no unequivocal definition of what counts as a "policy document". For the purpose of this project, we defined a *policy* as institutional-level documents or guidance that: 1) discusses, or aims to regulate, the use of (Generative) AI in education; 2) concerns direct educational uses (i.e., excluding all AI policies for marketing, branding, etc.); 3) addresses students and/or faculty (i.e., focuses on the interaction between GenAI and teaching and learning); and 4) have an institution-wide scope of application.

We analysed university policies using two approaches concurrently: a criterion-based analysis and documentary analysis. We obtained the policy via convenience sampling [95, 117], starting with our working group members which has representation from Australia, Canada, Jordan, New Zealand, Uganda, Italy, the United Kingdom, and the United States. We initially asked each member to gather their own university's GenAI policies, then asked each member to submit additional policy examples from their region. We then verified that we had adequate representation across geographical areas: EU, US/CA, Latin America, UK, AU/NZ, Arab Countries, Sub-Saharan Africa, Japan, China, India, South-East Asia. For the geographical areas that were under/not represented, we asked working group members to send policies from contacts/institutions in Asia, northern Africa, and the Middle East. We were not able to obtain policies for some of the universities we reached out to, or otherwise investigated, including one in the Netherlands and one in Morocco.

*3.3.1 Criterion-based analysis.* In policies that we did obtain, we ensured we had examples representing: 1) range of policies at different levels (i.e., whether they are general recommendations, implementation-ready policies, or in-between); 2) at least one non-PhD-granting institution; 3) both public and private institutions. For this round of analysis, we evaluated a total of 21 policies across a set of potential issues related to GenAI using a qualitative "stop-light evaluation system" (as examples of such systems, see [171] or [197]) where thorough addressing of a parameter was marked with a fully colored circle; implicit or partial addressing with a partially shaded circle; and not addressing the issue with an unshaded circle. The set of issues, listed below, was derived inductively by accretion. During the first read of the policies, we coded the ethical issues that policies discussed, defining one code per issue. We then proceeded to evaluate at what level of depth each policy addressed each ethical issue and what level of guidance they gave (if any) in defining actionable recommendations for addressing concerns in a given ethical issue.

- *Definition of AI*: Does the policy explain what they mean by "AI" in the context of their documents?
- *Grounding concepts*: Does the policy provide baseline information on how AI works, why it's important to discuss it in the context of teaching and learning, etc.?
- *Privacy*: Does the policy discuss protection of individual or collective privacy?
- *Bias*: Does the policy discuss how AI output may be "biased" (even without defining accurately what bias may be)?
- *Equity*: Does the policy discuss how AI output discriminates against given social groups?
- *Environmental sustainability*: Does the policy discuss the environmental costs and geopolitical consequences of developing and using AI?
- *Academic integrity*: Does the policy discuss how using AI may constitute academic misconduct, or a breach of academic integrity?
- *Plagiarism*: Does the policy discuss how using AI may lead to plagiarism?
- *Intellectual property*: Does the policy discuss issues of intellectual property, e.g., when asking AI to emulate the style of a given person?
- *Information accuracy*: Does the policy state that AI results may be inaccurate?
- *Consent*: Does the policy require people using AI to gather consent for its use in collective contexts (e.g., one student tasked with revising a group report using AI to revise other students' drafts)?
- *Impact on learning*: Does the policy discuss how AI may (negatively) impact student learning?
- *Transparency/disclosure*: Does the policy recommend people using AI to disclose its use, and how AI was used?

*3.3.2 Qualitative Documentary Analysis.* We drew from policy analysis questions [58] to conduct our documentary analysis of the policies. Table 4, adapted from Cardno [58], contains the probing questions we used to guide our documentary analysis of policy context, text, and consequences. For the documentary analysis, we annotated the policy documents (performing a round of open coding), identified repeated themes, and extracted categories and themes. We then grouped and mapped these themes through critically probing the policy context, text, and consequence, following Henry et al. [140]. To analyze policy purpose, we identified keywords or phrases that referred to the purpose and context of the policy as well as any reference to its underlying values. To analyze policy text, we probed for indications for who was responsible for developing the policy and how it was developed, and any alignment with local or national regulatory requirements. To analyze policy consequences, we identified words or phrases that reference the implementation or impact of the policy, including its strengths, weaknesses, opportunities, and challenges. A guiding principle we followed when conducting qualitative analysis of the policies was to identify what is stated in the policy and what is not, but can be inferred based on a holistic analysis of the document.

**Table 4: The probing questions (adapted from [58]) we used to interrogate policies in their content, text, and consequences.**

| Dimension of Analysis | Probing Questions |
|---|---|
| Policy context | What is the purpose of this institutional policy?<br>Is there evidence of drivers behind the policy?<br>What foundational values guide the policy?<br>Are there competing values at play that may create tensions? |
| Policy text | How does the text of the policy provide evidence of how it was developed?<br>Are there procedures outlined in the policy text that provide guidance for practice?<br>In what ways is the policy aligned with governance requirements? |
| Policy consequences | What is the intended impact of the policy?<br>How is the implementation of the policy to be monitored and reviewed?<br>In what ways does the policy text highlight crucial aspects of practice related to the policy? |

## 3.4 Key Findings

Table 5 summarises the policies we have analysed through our criterion-based analysis.

Since policy issues are grounded in values, policy scholars have shifted from the stance of policy analysis being an objective process to one that emphasizes transparency [138]. Values underlie how the problem is defined and solutions are identified [309]. The individual GenAI policy documents across institutions vary in terms of the explicit and implicit language used, which in turn reflects the community's values. While each policy inevitably reflects the context it was written for, it is still beneficial to discuss some of the key findings that emerged in the analysis of the policies and how they relate to the broader conceptual framework of the paper.

*Key Finding 1.* Based on our criterion-based and qualitative analysis, we propose a classification of the analysed policies situated along a linear continuum (see Fig. 7). On one end of this continuum are policies that are more "intellectual", or virtues-focused (i.e., outlining intellectual and moral principles for adoption of AI in education) [20]. Policies that are more "prescriptive" or compliance-focused (i.e., focus on abiding by the letter of the law, policies, procedures, and associated penalties for not doing so) are situated on the other end. Between these two ends are policies that offer guidance, "suggestions for effective use", or use-focused examples based on scenarios (e.g., referencing, summarisation, ...). This continuum of policies reflects a range of ethical stances from virtuous

**Table 5: A summary table of the criterion-based analysis for all the university policies.**

| | Institution facts | | | Policy overview | | | The policy discusses... / Provides actions to address... | | | | | | | | | | |
|---|---|---|---|---|---|---|---|---|---|---|---|---|---|---|---|---|---|
| ID | Region | PhD Granting | Public/Private | Detail level | Defines "AI" | Grounding concepts | Privacy | Bias | Equity | Sustainability | Academic integrity | Plagiarism | Intellectual property | Information accuracy | Consent | Impacts on learning | Disclosure of use |
| 1 | EU | Yes | Public | General | No | No | ●/◐ | ●/○ | ○ | ●/○ | ◐/○ | ◐/○ | ●/○ | ●/● | ○ | ○ | ○ |
| 2 | US / CA | Yes | Public | Implementation | Yes | Yes | ●/● | ○ | ●/◐ | ○ | ○ | ○ | ●/◐ | ●/◐ | ●/◐ | ●/● | ○ |
| 3 | US / CA | Yes | Public | General | No | No | ●/● | ●/● | ●/◐ | ○ | ●/○ | ●/● | ○ | ○ | ○ | ○ | ●/○ |
| 4 | US / CA | Yes | Public | Implementation | No | No | ◐/◐ | ○ | ○ | ○ | ●/● | ○ | ○ | ◐/◐ | ○ | ●/● | ○ |
| 5 | US / CA | Yes | Public | General | Yes | No | ●/○ | ●/○ | ●/○ | ○ | ○ | ○ | ●/◐ | ●/◐ | ○ | ○ | ●/● |
| 6 | US / CA | Yes | Public | Implementation | No | No | ○ | ○ | ○ | ○ | ●/● | ●/● | ○ | ○ | ○ | ○ | ○ |
| 7 | UK | Yes | Public | Implementation | Partially | Yes | ○ | ●/○ | ○ | ○ | ●/○ | ●/● | ●/○ | ●/○ | ○ | ●/● | ●/● |
| 8 | US / CA | Yes | Public | Implementation | Yes | Yes | ●/● | ●/● | ●/● | ●/○ | ●/● | ●/● | ●/● | ●/● | ●/● | ●/● | ●/● |
| 9 | US / CA | Yes | Public | In-between | Yes | Yes | ●/○ | ◐/○ | ◐/○ | ○ | ○ | ●/● | ●/○ | ●/○ | ○ | ○ | ●/○ |
| 10 | US / CA | Yes | Public | Implementation | Yes | Yes | ●/● | ●/● | ○ | ○ | ○ | ●/● | ○ | ●/● | ○ | ●/● | ●/● |
| 11 | AU | Yes | Public | Implementation | Yes | Yes | ●/● | ●/● | ○ | ○ | ●/○ | ●/● | ●/● | ●/● | ○ | ●/◐ | ●/● |
| 12 | US / CA | Yes | Public | In-between | Yes | Yes | ●/● | ●/○ | ○ | ○ | ●/○ | ●/● | ●/○ | ●/○ | ○ | ●/○ | ●/● |
| 13 | AU | Yes | Public | Implementation | No | No | ○ | ○ | ○ | ○ | ●/○ | ●/● | ○ | ●/○ | ○ | ●/◐ | ● |
| 14 | US / CA | Yes | Public | General | Yes | Partially | ○ | ●/○ | ●/○ | ○ | ●/● | ●/● | ○ | ●/○ | ○ | ●/◐ | ● |
| 15 | EU | Yes | Public | Implementation | Partially | No | ●/● | ○ | ○ | ●/○ | ○ | ●/● | ●/● | ○ | ○ | ○ | ●/● |
| 16 | CH | Yes | Public | Implementation | No | No | ○ | ○ | ○ | ○ | ◐/○ | ◐/○ | ○ | ○ | ○ | ●/○ | ●/● |
| 17 | CH | Yes | Public | Implementation | Yes | Yes | ●/● | ●/● | ○ | ○ | ●/● | ●/● | ●/● | ◐/◐ | ○ | ○ | ●/● |
| 18 | ME | No | Private | In-between | Yes | Yes | ●/○ | ●/○ | ●/○ | ○ | ○ | ●/○ | ○ | ○ | ○ | ○ | ●/○ |
| 19 | ME | Yes | Private | Implementation | No | Yes | ○ | ○ | ○ | ○ | ●/○ | ●/● | ○ | ●/● | ○ | ●/● | ●/○ |
| 20 | ME | Yes | Private | General | Yes | Yes | ○ | ◐/○ | ○ | ○ | ●/● | ●/● | ○ | ●/● | ○ | ●/○ | ○ |
| 21 | ME | Yes | Private | In-between | Partially | No | ○ | ●/○ | ●/○ | ●/○ | ●/● | ●/● | ○ | ○ | ○ | ●/○ | ●/○ |

● Thoroughly addresses issue ◐ Partially addresses issue ○ Issue is not addressed

When a policy does not discuss an issue, only a single ○ is present.

to utilitarian. It can be argued that the compliance-focused policies border more on legal than ethical in that they focus on not violating written policies.

It is important to note that while all university policies included elements of virtuous-lens, use-focused, and compliance-focused, our content analysis revealed that the relative balance of the ethical stances varied across policies. Each institution weighted compliance-focused, use-focused, and intellectual/moral/civic/principled/virtuous lens of their policies differently. Figure 8 portrays a set of conceptual examples that show varied weightings when approaching GenAI policy. Universities such as the University of Trento may adhere to more of a virtues-focused approach, discussing their vision on desirable uses of GenAI, and minimally specifying detailed actions that should be taken by individuals wishing to comply with the policy. Conversely, the University of Utah's policy weighs more of a compliance-focused approach, discussing how university personnel will address breaches of academic integrity due to misuse of GenAI. Other universities, such as Nanjing University, have policies that open with a virtue-focused approach, then shift to a structure of a legislative document. In contrast, the University of Waterloo outlines high-level principles, then maps them to legislation issued by the Government of Ontario.

We can also draw a yet closer look, by analysing individual statements. A virtue-based statement, taken from the University of Glasgow's AI policy, cites: "rather than seek to prohibit your use of these tools, we want to support you in learning how to use them effectively, ethically, critically, and transparently". Conversely, a compliance-based statement, taken from the University of Utah's policy, cites: "Plagiarism is a serious academic and research misconduct offense. The Research Misconduct Policy describes it as [omitted]. Full processes by which this is evaluated and addressed can be found at [link]". Statements may also include elements of both, and thus fall in between. Arizona State University's policy states: "Arizona State University is committed to the practice of Principled Innovation, embracing innovation with curiosity and wisdom (a virtues statement, ndr)...It is important to explore and respect university policy responsibly, protect your own privacy and the privacy of others, and keep in mind intellectual property considerations (a compliance statement, ndr)".

*Key Finding 2.* Institutional policies attempt to balance between the pace of rapid emergence of GenAI tools and capabilities and the comparatively slower pace of institutional change, including the adoption of policies. Many include clear statements that indicate that GenAI falls under the scope of the institution's existing technology system. University of Waterloo's guidance on GenAI, for example, states "All existing University policies still apply to the usage of AI, in the same way they do to all information systems." This guidance also includes the explicit recommendation to proceed "with caution and due diligence while using University data to train, prompt, or interact with AI systems that are not yet licensed or vetted," pointing to the emerging field of GenAI. While this institution attempts to 'future-proof' its GenAI policy and guidance by positioning GenAI as another "information system," it also responds in specific and direct ways in other contexts. For example, when describing the GenAI tools available through the institution, the policy document explicitly states that the University does not house the DeepSeek model and that it must not be used in conjunction with University data. University of Utah's policy includes a statement in large, bold letters within their "Prohibited Use of AI" section of the document that states, "This policy will be updated as new laws and legal review are enacted."

*Key Finding 3.* Policy context drives the institutional adoption of GenAI policy and guidance, resulting in a range of regional and institutional variations of policies. From the policy text, it is evident that values in each context play a key role in the policy development, implementation, and monitoring of their GenAI policies. Temple University, for example, states "As a university dedicated to exchanging ideas, distributing knowledge, and fostering innovation, we must embrace GenAI carefully and responsibly." University of Glasgow acknowledges from the outset that their students "will graduate into an AI-augmented world" and state that they have "a responsibility to prepare [them] for this world, providing space to experiment with, and understand the potential of, AI in an ethical way." In line with this value of preparing their students for the AI world, they state that they aim to support their students "in learning how to use [AI tools] effectively, ethically, critically, and transparently" rather than "prohibiting them from using these tools."

The socio-political context also plays a key role in institutional GenAI policies and guidance. Universities may reference national or local policy or legislation in their AI guidance. For example, the University of Waterloo explicitly states that the institutional use and development of AI aligns with that of the Government of Canada (i.e., fair, accountable, secure, transparent, educated, relevant) and the Government of Ontario (i.e., transparent and explainable, good and fair, safe, accountable and responsible, human-centric, sensible and appropriate). The University Mohammed VI Polytechnic's GenAI policy, when addressing the question of whether AI can be listed as an author, references the traditional understanding of Moroccan law, which "designates authors as natural persons who exercise skill and judgment in the creation of a work, thereby excluding AI entities from authorship."

## 3.5 Discussion

Among compliance-focused policies, the discussion is dominated by academic integrity, plagiarism, and IP concerns. This is not surprising, since IP and plagiarism concerns can be easily mapped to locally-applicable legislation (such as regional, national, or supranational laws on copyright and intellectual property), while issues of plagiarism and academic integrity can be mapped to pre-existing university policies on the same topics. When it comes to these matters of legal compliance, the advent of Generative AI in education often implies a simple redefinition of the scope of application of pre-existing policies.

Virtues-focused policies, on the other hand, are more heterogeneous: among the themes they cover, we highlight that a few policies discuss the significant environmental impacts of using GenAI (a topic that has attracted substantial attention elsewhere in academic literature, see e.g., [241, 270]) and policies discussing issues of consent (that is, the idea that if a GenAI tool is used in a group setting, all participants in the group work should consent to the use of the tool). The rapid pace of adoption of GenAI tools, when it comes to similar concerns, have significantly exacerbated the many socio-technical issues researchers have connected to education technologies in general [160, 311] and algorithmic harms related to GenAI [317] more specifically.

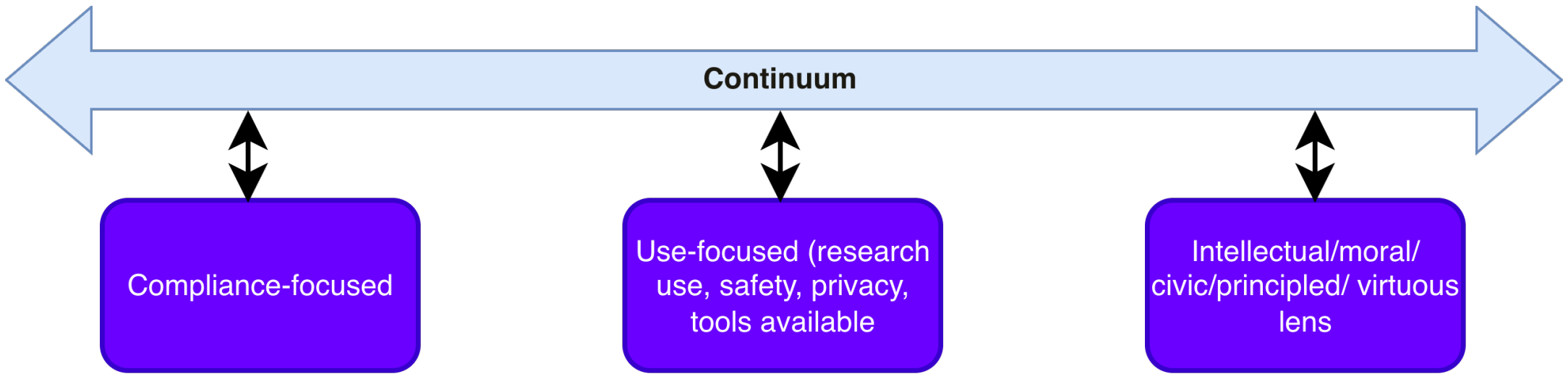

**Figure 7: Continuum of GenAI policies.**

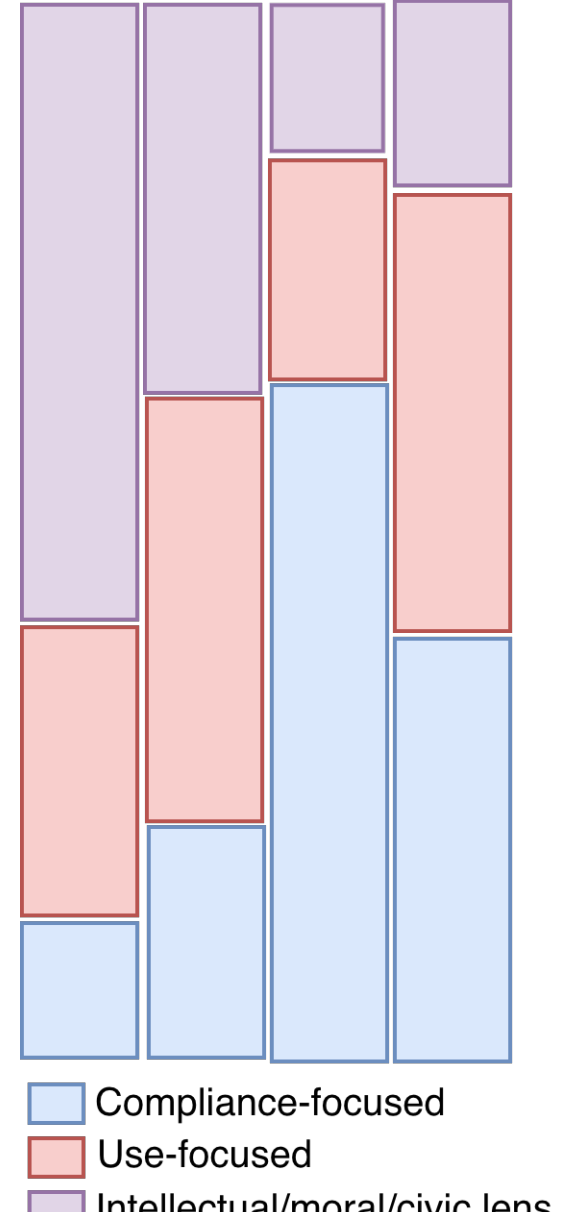

**Figure 8: Examples of varied weighting of institutional GenAI policies across compliance, use, and virtue domains.**

A common topic in many policies, independent from how they are positioned on the spectrum, is the discussion of potential impacts on student learning, data sharing implications, and its potential uses for teaching, learning and research (i.e., utilitarian lens). As we present in more detail elsewhere in this paper, one of the key tensions in the use of GenAI in education is in the potential adverse effects of cognitive offloading, especially in the students' development of writing skills, clear communication, and independent/critical thinking. We interpret the frequent presence of this topic in university policies as an indication that human-centered vs. machine agency remains indeed an unresolved tension with implications and impacts for policy and practice related to teaching and learning that are still emerging.

The key trends that emerge from the analysis of these policies inform the development of the ESI-Framework that we will now proceed to present.

# 4 The Ethical and Societal Impacts Framework (ESI-Framework)

In this section, we outline our method of deriving the ESI-framework, its elements, and how it can be used to navigate GenAI dilemmas in computer higher education. Existing conceptual analysis methods that involve merely determining the occurrence of concepts within texts lack the emphasis on understanding phenomena to be able to then generate theoretical insights [59]. Moreover, social phenomena, such as the ethical and societal impacts of GenAI, are complex and connected to multiple disciplines. Hence, we follow Jabareen's [154] conceptual framework analysis where we synthesize multidisciplinary literature through an iterative, qualitative, and grounded-theory approach, resulting in a network of linked concepts. Through this context-based, process-oriented approach [257], we develop each concept related to the ethical and societal impacts of GenAI and its specific function within the resultant framework [271].

## 4.1 Deriving the ESI-Framework

*4.1.1 Data.* The data sources selected to derive a conceptual framework should represent the various dimensions of the phenomenon we are exploring, namely, the ethical and societal impacts of GenAI in higher computing education. To this end, we leverage a range of data sources that include our working group's meeting notes, policy documents, articles, and guidance documents.

*4.1.2 Process.* We used an iterative and comparative process based on grounded theory [248] to derive the ESI-framework, continuously moving between and among our evidence, data sources and concepts to clearly define the scope and concepts.

*4.1.3 Procedure.* The method of deriving the ESI-Framework follows the eight-phase technique of Jabareen (2009) [154]: (1) Mapping data sources; (2) Extensive reading and categorizing the selected data; (3) Identifying and naming concepts; (4) Deconstructing and categorizing the concepts; (5) Integrating concepts; (6) Synthesising, resynthesising, and sense-making; (7) Validating the framework; (8) Rethinking the framework. In the following, we detail the work carried out in each of the phases 1 to 7. Phase 7 is covered in Section 4.2 while Phase 8 includes the implications of our results and future work ideas covered in Sections 4.3 and 6 respectively.

*Phases 1 and 2: Mapping data sources and extensive reading and categorizing the selected data.* We used the framework for implementing the social and ethical impact of computing developed by Martin et al. [227]. The literature review subgroup identified two additional topics of ethical analysis (Sect. 2.1.5) that we adopt to extend the Martin et al.'s framework: Computing Ethics and Professionalism (regarding the broader educational and professional focus of ethics) and Sustainability (regarding the sustainability of GenAI-based technologies). The literature review subgroup also provided a small, purposively selected set of papers ([66, 88, 123, 261, 279, 280, 374]) to which we added papers on a framework for reponsible AI education [38], lecturers' perspectives on the role of AI in personalized learning including benefits, challenges, and ethical considerations [246], and UNESCO's guidance for GenAI in education [236]. Based on the policies subgroup (Sect. 3.1), we integrated the Dabis & Csaki [73] ethical clusters of values and their data analysis of institutional policies related to GenAI.

*Phase 3: Identifying and naming concepts.* The framework for implementing the social and ethical impact of computing of Martin et al. [227] defines levels of social analysis and topics of ethical analysis. The levels of social analysis are: Individuals, Communities & Groups, Organizations, Cultures, Institutional Sectors, Nations, Global, while the topics of ethical analysis are: Quality of Life, Use of Power, Risks & Reliability, Property Rights, Privacy, Equity & Access, Honesty & Deception.

We adopt the definition of stakeholders of a computer system used by Gotterbarn and Rogerson [120] as individuals who are either significantly affected by or have material interests in the running of the computerised systems, and generalise our definition to include actors or groups who can affect or be affected by GenAI.

From Dabis & Csaki (2024) [73], we adopted the four ethical clusters of values: 1) Accountability and responsibility, 2) Human agency and oversight; 3) Transparency and explainability, and 4) Inclusiveness and diversity.

We define tension as a persistent conflict between different ethical values across all levels of social analysis (and implicitly between stakeholders or within the same group of stakeholders) based on their respective interests and demands [23, 220].

*Phase 4: Deconstructing and categorizing the concepts.* We revised the levels of social analysis identified by Martin et al. [227] for GenAI by identifying potential stakeholders or stakeholder characteristics for each level. For some pairs of level of social analysis and topic of ethical analysis, Martin et al. [227] identified hypothetical areas of concern for a particular technology resulting from the interaction. In addition, individual and professional responsibility are two additional lenses that should underpin the discussion of all possible areas of concern. In the following we identify (non-exhaustively) stakeholders for each level of social analysis:

- Individuals: student/learner (undergraduate and postgraduate students, lifelong learners, students with disabilities, underrepresented learners, international students, etc), educator/instructor (educator/instructor: lecturers, professors, tutors, teaching assistants, academic advisors, curriculum designers, etc.), computer professional, etc.
- Communities & Groups: student societies/unions, educator networks or associations, computer professional societies, etc.
- Organizations: AI-based platforms and infrastructures: tech corporations offering GenAI tools (e.g., OpenAI, Google, Microsoft), GenAI-based learning management systems providers, etc.
- Cultures: pedagogical cultures, professional identities, etc.
- Institutional Sectors: Higher Education institutions (senior management group, head of department, dean, university administrators, IT services, academic integrity panels, etc)
- Nations: governmental policies, accreditation agencies, organizations employing computer professionals (public or private sector), etc.
- Global: global governance bodies (e.g., UNESCO, OECD, European Commission), international computing bodies (e.g., ACM, IEEE), etc.

To generate an initial set of tension codes, we conducted an inductive coding process [336] on the set of key papers provided by the literature review subgroup. We openly identified specific codes for tensions as they arose from these papers. Some examples of codes include:

**T1:** *Academic freedom vs. centralized GenAI policies* as tension between institutional consistency and faculty autonomy. Stakeholders: Universities, Faculty. Ethical impact: Use of power [88].

**T2:** *Loss of foundational skills vs. efficiency using GenAI as tension between students possibly skipping essential learning stages and educators struggling to maintain educational depth.* Stakeholders: Students, Educators. Ethical impact: Impedes Learning [88].

**T3:** *GenAI-enabled inequality due to access gaps, inclusion and access equity vs. digital divide* as tension between GenAI risking widening educational disparities if access is unequal and GenAI requiring adequate technological infrastructure. Stakeholders: Students, especially Global South, Universities, Public. Ethical impact: Equity and access [236, 246].

**T4:** *Cultural dominance in GenAI vs. diversity and inclusion* and *Protection of cultural diversity vs. dominant AI training data biases* as tensions between marginalization of indigenous and non-Western knowledge systems. Stakeholders: Public, Minority Communities, Policymakers. Ethical impact: Bias, transparency and explainability [236].

**T4:** *GenAI's psychological effects vs. inclusive learning environments* as tension between possibility of harm to student well-being and affecting students' sense of inclusion. Stakeholder: Students. Ethical impact: Quality of life [38].

**T5:** *Increased efficiency via use of GenAI vs. job security for educators* as tension between the risk of displacing roles or devaluing educator expertise and creating fear of role displacement. Stakeholders: AI Industry, Universities, Educators. Ethical impact: Sustainability and global economy [88, 246].

*Phase 5: Integrating concepts.* Four authors, representing each of the three sub-groups (literature review, policy, and framework), engaged in an iterative process of synthesizing the themes and defining the scope of the concepts related to our phenomenon of GenAI in the higher education computing landscape. Through this process, we integrated concepts from each of our sub-group's data analysis to derive the ESI-Framework. We drew from the initial tension codes (see Phase 4), clusters of ethical values, topics of ethical analysis, policy data, and collective expertise and judgement of the researchers involved to identify crosscutting conceptual categories of tensions. While the initial codes used to derive the ESI-Framework was inductive and data-driven (see Phases 4), Phase 5 was informed by expert discussion, reflection, and synthesis reflected in meeting notes and transcripts. At the end of Phase 5, we drafted a diagram that depicts interacting themes, categories, and social levels at play when navigating the GenAI landscape in higher computing education.

*Phase 6: Synthesis, resynthesis, and making it all make sense.* Once we had an initial conceptualization of the landscape of GenAI tensions based on the diagram sketched at the end of Phase 5, we continued to meet with representation across groups, synthesizing the themes, categories, and tensions. Through this process we continued to iterate, synthesize, and theorize regarding the emerging relationships among the categories of tensions and topics that emerged from analysis of data on the ethical and societal impacts of GenAI. The interactions between tensions, categories, and clusters of ethical values can be seen in the ESI-Framework listed in Table 6. For example, the tension "Human-centered vs. Machine Agency" was mapped to the ethical cluster of "Accountability and Responsibility" and associated with the themes of honesty and deception, computing ethics and professionalism, and use of power.

To verify [238] our resulting framework, we matched the five examples of initial tension codes listed in Phase 4 against the categories of tensions in the ESI-Framework:

- **T1:** Academic freedom vs. centralized GenAI policies contributes to the topic theme Governance/control vs. Agency as it reflects the broader issue of who governs technology use in education and who retains decision-making agency.
- **T2:** Loss of foundational skills vs. efficiency using GenAI contributes to the topic theme Human-centered vs. Machine Agency as it questions whether the educational value remains centered on the learner's skill development or is ceded to GenAI tools or systems acting on behalf of the learner.
- **T3:** GenAI-enabled inequality due to access gaps, inclusion and access equity vs. digital divide contributes to the topic theme Human-centered vs. Machine Agency as it reflects whether all humans are equally positioned to benefit from machine agency or if human capacity is undermined by structural disparities.
- **T4:** Cultural dominance in GenAI vs. diversity and inclusion and Protection of cultural diversity vs. dominant AI training data biases contribute to the topic theme Justice vs. Exclusion as it reflects a justice-oriented concern with representational fairness and inclusion vs. the exclusion of various cultural perspectives in knowledge production and educational content.
- **T5:** Increased efficiency via use of GenAI vs. job security for educators contributes to the topic Fairness vs. Bias as it reflects an ethical bias towards valuing technological performance over human labor.

We also matched the ESI-Framework themes and categories against the codes that emerged from the literature review sub-group. In so doing, we were constructing our theory as our process of iterative analysis proceeded.

As part of Phase 6, we resynthesized the ESI-Framework by asking ourselves: *What are the implications of the interactions among the tensions, ethical values, and societal impacts identified in this framework?* To this end, we constructed a set of reflective questions to deconstruct each tension and support stakeholders as they navigate the complex and nuanced landscape of GenAI in higher education. These metacognitive questions are a core component of the ESI-Framework and reflect the iterative and comparative cycles of data collection, analysis, synthesis, and resynthesis across our three sub-groups. For example, the questions that correspond with the tension of "Human Centered vs. Machine Agency" include: *How are roles of humans and AI conceptualized? Is human judgment being deferred to machines? Are hybrid agency models supported and critiqued?* While not intended to be an exhaustive list of questions, they are intended to ignite further questions and to more deeply probe the tension when making decisions around GenAI. The questions that correspond with each tension, therefore, are central to our intentional attempt to answer the question: *How might the ESI-Framework be used by any stakeholder engaged in decision-making around GenAI usage, development, and deployment?*

*Phase 7: Validating the ESI-Framework.* Validating the conceptual framework involves addressing the central question of whether researchers, practitioners, and other stakeholders can make sense of the framework and apply it to their respective domains. In our case, the ESI-Framework should be applicable to stakeholders/actors at all social levels identified during Phase 4. In Sect.4.2 we offer dilemma analysis as an approach to validate the ESI-Framework.

*Phase 8: Rethinking the ESI-Framework.* Given the complexity and highly nuanced landscape of tensions related to the emerging and rapidly changing field of GenAI, the ESI-Framework can and should be revisited and revised based on new developments, research, insights, and documented benefits or harms (see Sect. 6).

**Table 6: The Ethical and Societal Impact (ESI) Framework**

| Cluster of ethical values | Topics of ethical analysis | Tension | Guiding questions |
|---|---|---|---|
| Accountability & responsibility | Honesty & deception<br>Computing Ethics & professionalism<br>Use of power<br>Proprietary Rights | Rights vs. Responsibility | What rights vs. responsibilities do individual users of AI have? What rights vs. responsibilities do policymakers have? |
| | | Human-centered vs. Machine Agency | How are roles of humans and AI conceptualized? Is human judgment being deferred to machines? Are hybrid agency models supported and critiqued? |
| | | Imagination vs. Inevitability | What futures are being imagined through this work? Are speculative or visionary practices encouraged? How is collective dreaming scaffolded? |
| | | Pace of change vs. Constraints of time | What lags exist between AI, ethics, and institutional readiness? How are ethical foresight and responsiveness being developed? Where are we disrupting without deliberation? |
| | | Hidden vs. Visible Labor and Value | Whose labor is visible and whose is hidden? How is value assigned to maintenance, care, and community roles? Are new forms of value being recognized? |
| Human agency & oversight | Quality of life<br>Risks and reliability<br>Computing ethics and professionalism<br>Use of power | Governance/control vs. Agency | Who is setting the agenda for AI adoption in this context? What power dynamics influence infrastructure decisions? How are community voices involved in governance? |
| | | Emotional and Affective Cost vs. Benefit | How is emotional labor recognised in AI-supported systems? What psychological impacts are emerging? Are care and relationality factored into design? |
| | | Compliance vs. Agency | Where is refusal or critique emerging? How is dissent supported or suppressed? What alternative pathways are being imagined? |
| | | Futures of work/education vs. Current reality | How are we preparing for uncertainty and/or for the current state of GenAI adoption in industry? What futures are we actively designing toward? How do we equip learners and workers for roles that don't yet exist? |
| | | Individual vs. Collective/Societal Good | What individuals oversee the output from the GenAI models? Should the oversight be done by an individual or a collective group of individuals? |
| Transparency & explainability | Risks and reliability<br>Computing ethics and professionalism<br>Privacy | Trust and Transparency vs. Opaqueness | What is the basis for trust in AI systems here? How transparent are the models, data, and design processes? Are learners equipped to interrogate black box systems? |
| | | Responsiveness vs. Stability | How is change perceived and managed? Are feedback loops in place and trusted? What adaptive capacities are emerging? |
| Inclusiveness & diversity | Equity and access<br>Quality of life<br>Sustainability | Access vs. Barrier | Who has access? What barriers exist? To whom and in what context? |
| | | Ecological Cost vs. Benefits | What is the environmental cost of this technology? Are planetary boundaries considered in decision-making? How is sustainability integrated into tool development, deployment, and scaling practices? |
| | | Justice vs. Exclusion | Who is included or excluded by design? What systemic inequities are reinforced or disrupted? How are equity and justice made visible and accountable? |
| | | Fairness vs. Bias | What systemic inequities are reinforced or disrupted? |

## 4.2 Validating the ESI-Framework through Dilemma Analysis

Dilemma analysis is a methodological approach to analyzing tensions and decision-making in organizational contexts [359]. The approach developed as a specific strategy to help bridge the theory-practice gap in educational action research. Winter [359] explained that "as a teaching practice (T.P.) supervisor/researcher", he sought to transcend his "view as a supervisor to create an account of the T.P. situation which would be faithful to the views of students, classroom teachers, and pupils, as well as those of fellow supervisors. This account had to gain the assent of all parties so that it could be used to illuminate for each party the point of view of the others, as a practical contribution to preparation for T.P.". The action-research task then, was to create an account of a situation which would be seen by a variety of others as convincing, i.e., as "valid".

Building on that work, McKernan [231] proposed an empirical approach, using the formal theory of contradiction to guide dilemma analysis: "i.e. that institutions have conflicts of interests, that members are split and divided, and all of this is beset by dilemmas" . The classic procedures for dilemma analysis involve conducting interviews, then analyzing the data in terms of a number of dilemmas, tensions or contradictions (categorized as ambiguities, judgments and problems) [69]. A set of perspective documents can be developed, organized by these dilemma categories from the perspective of each of the actors. These can then be used to create a profile perspective for each role in the setting.

So while the origins of dilemma analysis may lie in educational action research [231, 359], to the authors' knowledge it has not been previously applied in computing education research. Most closely to a computing context perhaps, it has been used in software engineering research as a critical method to unpack the dilemmas arising in a global software engineering context of a failed software project-*The Novopay Project* a nationwide New Zealand teachers' payroll replacement [69]. An example of one of the dilemmas identified sees the tensions relating to the vendor's dilemma of ***How to better understand customer context*** with one pole seeing a global hiring strategy being adopted *Hire/sub-contract nearshore or offshore* and at the alternative pole the vendor adopting a local hiring strategy *Hire/sub-contract 'in-country'*. A further dilemma example (with direct impact for the schools employees) covered how administrators when coping with ***Handling delay from service centers*** elected to *Work late (extra hours, miss family time)* or *Go home on time.*

While dilemma analysis is leveraged here as a method to validate the ESI framework, it is acknowledged that ACM does have an existing process guide for computing professionals and ethical practice known as *Proactive Care*. They advise that Proactive CARE uses the ACM Code of Ethics and Professional Conduct as a framework to identify and address opportunities to engage in more ethical computing practices [254].

That guide has four steps:

- CONSIDER who might be affected and how:
- ANALYZE the situation's details:
- REVIEW other obligations and limitations:
- EVALUATE the best course of action:

So, while *Proactive Care* offers a useful and concrete mechanism for reflecting upon the ethical dimensions of a specific context, the ESI framework promoted here extends beyond the perspective of an individual computing professional. We now elaborate on how dilemma analysis may be applied to validate the ESI framework.

*4.2.1 Unpacking Dilemmas.* In this study, we adopted a form of dilemma analysis based on the ESI-Framework (see Table 6) and grounded in the policy and literature data sources identified in this working group report. The dilemmas we considered are closely related to the tensions defined previously. Dilemmas presume a scenario/context and a tension/decision point that straddles the context with the ethical or practical consequences. Dilemmas and their associated tensions can further be mapped to different stakeholders which represent different perspectives. We developed the notion of *Dilemma Archetypes* to shift from focusing ethical dilemmas or tensions as individual acts to viewing them as a byproduct of systemic interactions [69]. Three exemplar archetypes were derived based on: empirical insights from the literature; abstraction from the origins and early tensions of AI as a technology; or mapping to one of the four clusters of ethical values. These archetypes were unpacked into subordinate dilemma sets for each archetype. These subordinate sets of dilemmas captured the perspectives of each role involved in the dilemma applicable for their role drawn from the archetype. The roles were based on the social levels identified by Martin et al. (1996) [227]. The mapping process conducted here constitutes a simulation-based form of validation of the ESI framework. Table 7 describes the three exemplar archetypes DA1, DA2, and DA3, highlighting the main description of each of the dilemmas.

We analyze these dilemmas using an adapted version of Bacchi's [28] "What's the Problem Represented to be?" (WPR) approach. This methodological strategy enabled us to critically interrogate the ethical dilemmas posed by the introduction of GenAI into higher computing education, the tensions these dilemmas encapsulate, and how stakeholders are affected by different choices. We modify the WPR approach [28] and pose the following two questions in our dilemma analysis:

**Q1:** What is the nature of the dilemma?
**Q2:** What alternatives should be acted upon to resolve the dilemma?

Question Q1 involves identifying the problem or dilemma being represented, how this dilemma came about, and the presuppositions or assumptions latent within the dilemma. This question also asks us to consider what remains as problematic in this dilemma and whether this dilemma can be thought about using an alternative framing. Question Q2 asks us to consider how this dilemma has been produced, how it could be disrupted, and the possible effects that are produced or amplified as a result of this dilemma.

*4.2.2 Dilemma Archetype DA1: Complete Cognitive Offloading vs. Conscious Critical Partnership.* The first dilemma archetype concerns tensions related to the use of GenAI and what the aims associated with the use of GenAI should be. Crucially, in this first dilemma archetype, the actor makes the choice and is choosing for themselves.

**Table 7: Dilemma Analysis - Archetypes**

| Left hand pole | Dilemma Archetypes (DA) | Right hand pole |
|---|---|---|
| Complete cognitive offloading | **[DA1]** Should the individual approaching a task aim to offload the task as much as possible to a GenAI system, or should they consciously and critically evaluate why, when, and how they use the GenAI system? ⟵⟶ | Conscious critical partnership |
| Complete automation with AI | **[DA2]** Should decision-makers seek to replace human efforts with AI systems, or should they only use AI to augment existing capabilities? ⟵⟶ | Augmentation of human effort |
| Non-disclosure of use | **[DA3]** Should an individual working on a task be required to disclose AI use, or should the use of GenAI be treated like any other tool (that is, without needing to specify whether it was used)? ⟵⟶ | Complete detailed disclosure |

> Should the individual approaching a task aim at offloading the task as much as possible to a GenAI system, or should they consciously and critically evaluate why, when, and how they use the GenAI system?

This archetype encapsulates the dilemma of the opportunity cost between reappropriating one's own time (plausibly, to be used in other activities unrelated to the task at hand) and relinquishing the learning that would be gained by performing the task, or vice versa (i.e., using GenAI tools to maximise learning, at the cost of still investing a significant amount of time). This maps to the tension of Human-centered vs. Machine Agency in the ESI-Framework where the guiding questions can support stakeholders with analyzing the nuances within this dilemma: How are roles of humans and AI conceptualized? Is human judgment being deferred to machines? Are hybrid agency models supported and critiqued?

The archetype is intimately connected to the second one (detailed below), which is connected to the "technological unemployment"/"technological replacement" narrative [189]. In this narrative, advanced machinery (thus including GenAI) should enable actors to decrease the amount of time they spend on completing a given task. In spaces of learning or work, however, do we have any guarantees that time "saved" by employing advanced machinery will be used to perform more productive tasks that are still conducive to the development of the learners'/workers' needed competencies?

Actors facing this dilemma should decide how much they are willing to invest in the competencies by performing the task: what competencies would they gain by adopting a "critical partnership" model (i.e., inquiring the why, when and how to use a given technology)? What would they gain or lose by increasing cognitive offloading? The positioning of an actor on the continuum between Complete Cognitive Offloading (CCO) and Conscious Critical Partnership (CCP) should reflect the outcome of their decision. Investment in competency development would bring them towards CCP; potential for enhanced productivity by cognitive offloading would bring them towards CCO. The CCO/CCP dilemma is actually highly nuanced and only a partial view of the larger issue: the question is not only of an actor's positioning in a continuum between CCP and CCO, but also their agency in deciding whether to use or reject the tool and the extent to which they use it. This dilemma also raises questions about paradoxical impacts of AI (e.g., may hinder productivity by creating new inefficiencies) and ethical questions (e.g., who is responsible for outputs in the case of CCO?) [321].

In the following, we discuss possible instantiations of this dilemma archetype DA1 for different classes of actors.

*Individual stakeholder - students.* Students may be tempted by trying to offload "boring homework" to a GenAI tool — but if they do so, will they still learn? Their dilemma would be:

> As a **student**, do I try to use GenAI tools to do coursework/exams in my stead (complete cognitive offloading), or do I try to use them to enhance my own capabilities (conscious critical partnership)?

*Q1: What is the nature of the dilemma?* The dilemma hearkens back to a debate on the nature of using tools as aids for learning that extends as far back as ancient Greece, with dialogues such as the Phoedrus discussing the nature of writing. The question always is: What human capabilities are we relinquishing, if any, by using tools as cognitive aids?

*Q2: What alternatives should be acted upon to resolve the dilemma?* In finding a resolution for the dilemma, students should consider not only how they use the tool (and thus, positioning themselves more towards CCO or CCP), but also whether they should use GenAI at all. Renouncing partnership is, in some way, a form of conscious critical partnership.

*Individual stakeholder - educator.* They may be tempted to offload some of their own cognitive labour to "free up time". What are, however, the risks that emerge from this? We propose the following dilemma:

> As an **educator**, do I use GenAI to grade my students' work, or do I use it — if I feel the need for it — to check whether I fully addressed the feedback rubric/checklist?

**Table 8: Dilemma sets for dilemma archetype DA1: Offloading vs. Partnership.**

| **Left-hand pole**<br>**OFFLOADING** | **Dilemma Sets**<br>⟵⟶ | **Right-hand pole**<br>**PARTNERSHIP** |
|---|---|---|
| Delegate exams and coursework to GenAI | As a **student**, do I try to use GenAI tools to do coursework/exams in my stead (complete cognitive offloading), or do I try to use them to enhance my own capabilities (conscious critical partnership)? | Use GenAI to enhance my own capabilities |
| Grade students in my stead | As an **educator**, do I use GenAI to grade my students' work, or do I use it — if I feel the need for it — to check whether I fully addressed the feedback rubric/checklist? | Check tone/validity of feedback |
| Complete vibe coding | As a **software developer (i.e., a computing professional)**, I can choose whether I want to use GenAI to fully "vibe code" a given programming task, or to use it for less critical tasks (e.g., providing first examples of non-critical tests in a test suite). | Validating test suites |
| Generation of study plans | As a **university (i.e., an organisation)**, we can use GenAI to produce drafts of degree study plans, or curate our own study plans, using GenAI to suggest potential gaps in coverage of competencies. | Highlight missing competencies in study plans |

*Q1: What is the nature of the dilemma?* Educators may be tempted to lean towards CCO to "free up their time", but how should the freed-up time be used? Time used in grading can, in some way, represent time spent in caring for students. Educators should also consider whether time spent grading represents — in some way — an opportunity for an educator's own professional development.

*Q2: What alternatives should be acted upon to resolve the dilemma?* Educators who do not have a clear vision about how to use the freed-up time for care should explore what opportunities for additional care, or for additional learning, they see for their students. Resolving the dilemma is not only about leaning towards CCO or CCP, but also about deciding how one's energies are directed and invested when it comes to duties of education and care towards students.

*Individual stakeholder - professional.* The dilemma we propose for professionals is that of a software developer deciding to use GenAI to perform tasks connected to their mainline job:

> As a **software developer**, should I use GenAI to fully "vibe code" a given programming task, or should I use it for less critical tasks (e.g., validating examples of non-critical tests in a test suite)?

*Q1: What is the nature of the dilemma?* There has been ample discussion on the effectiveness, solidity, and security of code produced by GenAI tools. Errors that GenAI tools may make not only in the code itself, but also in the code's documentation, potentially compromise the stability and success of entire software projects. "Vibe coding" may be a good way to close a task fast, but what could be the long-term costs?

*Q2: What alternatives should be acted upon to resolve the dilemma?* Developers should question what cognitive tasks can be safely offloaded to GenAI, and what — if any — opportunities GenAI may offer to form multi-actor (e.g., developer: generative tool) partnerships. Much like in the previous case, software developers should also position themselves on the continuum between high and low use of GenAI.

*Collective stakeholder - organisation.* We also propose a dilemma formulation from the perspective of a university department. In fast-changing disciplines such as computing, a university may be tempted to completely offload the task of generating a study plan to a GenAI tool. The question, however, is whether the tool would be able to effectively capture the nuances and needs of a study plan.

> As a **university**, should we use GenAI to produce drafts of degree study plans, or curate our own study plans, using GenAI to suggest potential gaps in coverage of competencies?

*Q1: What is the nature of the dilemma?* A study plan is not just a collection of courses, but courses also need to fit specific institutional criteria, which the GenAI tool may not take in consideration. Additionally, if a field is subject to fast change, the inductive nature of GenAI tools may lead them not to duly emphasise the specific challenges that future graduates will need to face when working in that discipline.

*Q2: What alternatives should be acted upon to resolve the dilemma?* Since GenAI tools can perform tasks of summarisation and rough pattern-matching, universities could look at uses of GenAI that, for example, check study plans for potential saturation or spaces of opportunity.

*4.2.3 Dilemma Archetype DA2: Complete automation with AI vs. Augmentation of human effort.* Should decision-makers seek to replace human efforts with AI systems, or should they only use AI to augment existing capabilities? This dilemma is the prescriptive version of the CCO-CCP dilemma, in cases where relevant actors are empowered to make decisions that direct or determine the roles and actions of others. Fundamentally, the decision maker must determine whether some individual or group should employ AI (or not) to accomplish some task or perform an action.

The rationale for automation may include reduced cost, time savings, greater precision, etc. Reasons for favoring more limited augmentation might be fairness, authenticity, social or cultural mores, workers' rights, etc. Or, as Denning has framed it, *"A definition of productivity that prioritizes speed of task completion over amplification of human capabilities - AI replacing human work rather than augmenting it"* [79].

The consequences of decisions captured by this dilemma may be associated with individuals, definable groups, or society at large. The nature and identities of the decision makers and those affected by their choices determine which tensions are in play for a given instantiation of this dilemma. At the center of this navigating this dilemma is the "Human agency & oversight" cluster of ethical values in the ESI-Framework. This leads stakeholders to pose questions related to risks and reliability, computing ethics and professionalism and use of power such as: Who is setting the agenda for AI adoption in this context? What futures are we actively designing toward?

For example, when deciding on AI usage policies governing course assignments, educators must navigate the tensions between students desire to complete assignments quickly and efficiently and learning outcomes associated with assignments [331]. Instructors must determine whether using AI to accomplish an assignment runs afoul of the learning outcomes associated with that assignment. If using AI does not interfere with meeting learning goals, then use of such tools might be permitted; otherwise, instructors might seek to limit usage.

In a professional context, decisions regarding AI-enhanced software development involve tensions between worker rights, happiness, and professional reputation and organizational productivity, quality, and efficiency. Organizations that produce software must balance the time saved by using automated coding tools (e.g., ChatGPT, Copilot) with other considerations [101]. In some contexts, there are business cases for limiting AI usage (e.g., correctness, security, IP protection). In addition, there may also be moral and/or philosophical reasons for limiting AI deployment related to the nature of work and the role of human producers in society.

Professional communities must also grapple with how AI gets included in shared activities. For instance, for many academic publishing venues, the number of submitted papers outpaces the size of the reviewing workforce, putting reviewer time and review quality in tension [144]. Reviewers who find themselves with excessive workloads might prefer to save time by leveraging AI systems to summarize and critique papers that they have been assigned to review [1]. However, delegating peer review to a GenAI system might compromise the usefulness and quality of generated reviews.

*4.2.4 Dilemma Archetype DA3: Non-disclosure of use vs. Complete detailed disclosure.* This archetype gathers the dilemma of an individual working on a task, trading off non-disclosure of the use of GenAI in their work at the cost of providing complete and detailed disclosure and acknowledgment of the use of AI. In doing so, they either disclose their use of GenAI tools with transparency, or treat the use of GenAI tools as any other tool where one does not need to acknowledge its use. This archetype presents an ethical dilemma to transparently disclose the use of GenAI tools or maintain a level of opacity. This framing neither precludes nor assumes a net productivity benefit with the use of GenAI for either the organisation or the individual.

The positioning of an individual on the continuum between non-disclosure of use and complete detailed disclosure should reflect the outcome of their decision. Investment in basic knowledge of AI use, and understanding ways that allows others to see clearly why, what, and how it is being done would bring them toward the right end; non-disclosure of AI use would bring them toward the left end.

This dilemma is nuanced and highly contextual, lending only a partial view of the larger issue: the question is not only of one's positioning in a continuum between non-disclosure versus complete detailed disclosure, but also between the communication, openness, transparency, trust, and accountability related to the use/non-use of GenAI tools. Analyzing this dilemma centers on the ESI-Framework's cluster of ethical values of "Accountability & responsibility" and the topics of "Honesty & deception" and "Computing Ethics & professionalism." This dilemma also relates to the "Transparency & Explainability" in the ESI-Framework and stakeholders can pose questions to support dilemma analysis: Are feedback loops in place and trusted? How is value assigned to maintenance, care, and community roles? Most of the university policies that were reviewed in Section 3.4 require that students and educators disclose the use of GenAI tools in their assignments or work. However, this creates a dilemma: the loss of convenience and agency that comes with the disclosure of use requirement when using the tools vs. the loss of credibility if there is non-disclosure.

This dilemma archetype presents a tension across several stakeholders, including students, teachers, or even professionals who use GenAI tools publicly and privately. From a social perspective, recent work shows that students' use of AI and social norms are shaped by their peers [145, 146, 262]. Consequently, the willingness to disclose AI usage is also heavily influenced by students' perceptions of whether and how their peers disclose their use. In a study by Hou et al. [145], students described 'hiding' their AI use and felt shameful about using the tools publicly in front of their peers. So, while instructors can and do set policies that influence student behavior, these norms also have the power to shape behaviors and to reify emerging norms around public or private use of GenAI in the classroom.

In many cases, the use of GenAI is difficult to acknowledge because it could be used to brainstorm ideas, debug code or even refine one's writing, albeit making it hard to clearly delineate where one's own contribution ends and where the AI's contribution begins. Recent studies have even demonstrated that students are often unaware of the extent to which they rely on AI support [282], blurring the lines of attribution. There is lack of consensus about best practices of the use of AI in the peer review process with initial research advocating for a gradual approach with close human oversight to maintain and uphold trust, transparency, and research quality [310].

## 4.3 Discussion

Navigating the age of GenAI is complex and nuanced, as evidenced by the three dilemma archetypes. Applying the ESI-Framework, we demonstrate how to deconstruct dilemmas and generate alternative paths of action to arrive at constructive ends whether related to learning outcomes or greater social good. Through evaluating the coherence and practicality of the ESI-framework via three illustrative dilemma archetypes, we used each of the dilemma archetype to test the capacity of supporting multi-stakeholder reasoning. Dilemma analysis also allowed us to simulate how the framework could be applied to specific scenarios of ethical challenges in GenAI use within computing education.

**Table 9: Dilemma set for dilemma archetype DA2: Automate vs. Augment.**

| **Left-hand pole**<br>**AUTOMATE** | **Dilemma Sets**<br>⟵⟶ | **Right-hand pole**<br>**AUGMENT** |
|---|---|---|
| Allow full use of GenAI to complete the assignment | As an **instructor**, do I permit (or encourage) my students to utilize GenAI tools for completing writing assignments (automation), or only allow them to correct for spelling and grammar (augmentation)? | Prohibit use of GenAI |
| Allow full use of GenAI to generate proprietary code | As a **manager** of a software group, do I encourage my team to increase productivity by using code-generation tools to enhance proprietary software (thus exposing codebase to third parties, potentially introducing security threats, etc.), or prohibit their use? | Prohibit use of GenAI |
| Allow AI-generated paper reviews | As an academic and professional **interest group**, do we allow peer-reviewers to offload the review of submitted conference papers to AI systems, or do we disallow their use when reviewing? | Prohibit use of GenAI when reviewing submissions |

**Table 10: Dilemma set for dilemma archetype DA3: Disclosure vs. Non-disclosure.**

| **Left-hand pole**<br>**NON-DISCLOSURE** | **Dilemma Sets**<br>⟵⟶ | **Right-hand pole**<br>**DISCLOSURE** |
|---|---|---|
| Non-disclosure of use of GenAI to complete an assignment | As a **student**, working on a task should I disclose or provide complete, detailed acknowledgment of GenAI use in my work or do I present the work solely as my own, even if it blurs the lines of individual contribution? | Full disclosure of use of GenAI to complete an assignment |
| Non-disclosure of use of GenAI to complete assignment | As an **instructor**, do I permit (or encourage) my students to submit their work without a complete and detailed acknowledgment of the use GenAI tools in their assignments (automation) or not? | Full disclosure of use of GenAI in the assignment |
| Non-disclosure of use of GenAI in policy/guideline | As a **university administrator**, do I focus my policy/guidelines on the full disclosure of the use of GenAI for example, for student assessment, or encourage) permit the use of GenAI tools as any other tools that do not need disclosure? | Full disclosure of use of GenAI in policy/guideline |
| Non-disclosure of the use of GenAI to complete peer reviews | As an academic and professional **interest group**, do we allow peer-reviewers to submit reviews without complete and detailed disclosure of the use of GenAI tools in the review process, or do we view GenAI tools as any other tools that people can use and therefore do not need to acknowledge? | Full disclosure of use of GenAI to complete peer reviews |

While we designed the ESI-Framework to guide ethical reflection on concrete cases, such as learning design, integration of GenAI tools in assessment, or institutional policy, this approach of pairing the ESI-Framework with dilemma analysis can be extended to additional dilemmas related to the ethical and societal impacts of computing. Users of the ESI-Framework can begin with a specific dilemma or scenario and identify the primary stakeholders (or perspectives) and social level(s). They can then use the ESI-framework to identify which the ethical values most at stake, topics of ethical analysis that are implicated, and then use the guiding questions provided as a starting point to structure the ethical reflection and deliberation. Users can then repeat the ESI-Framework process for different stakeholder perspectives and identify how and where they might interact. Thus, we recommend applying the ESI-Framework as a flexible lens from the perspective of different stakeholders who are involved or impacted by the given scenario as this will illuminate possible ethical vulnerabilities, trade-offs, or benefits.

Central to these dilemmas are questions such as: How can I act ethically and in ways that acknowledge and honour the needs of all stakeholders? How can I ensure that the principles and guidelines of the professional societies of which I am a member (e.g., ACM and IEEE/ACM codes of ethics) are carefully adhered to when developing or purchasing AI systems and Educational Technology platforms? How can I embed ethics, mitigate sociotechnical harms through harm-envisioning tools, and design GenAI with equity at the forefront? How can I inspire and integrate this practice, mindset, and approach into designing and teaching computing courses?

## 5 Limitations & Assumptions

The systematic literature review checked three key databases (ACM, IEEE and Scopus) which we maintain provided a sufficiently broad coverage of peer reviewed literature on the the topic of ethical and societal impacts of GenAI in computing higher education.

However, this is a rapidly evolving field and is characterized by highly active research into an emerging technology. This interest was evidenced from our paper filtering process by the initial 3827 references returned from our initial searches. Thus, continuing developments, including those reported in non-refereed publications and grey literature sources, will not be incorporated. Nonetheless some references to arXiv [101, 365] have been incorporated in the wider report. Yet we believe we have captured a comprehensive landscape representing the field of computing higher education and GenAI as reported at the time of the review.

Given the volume of papers and time available, we also chose to limit the scope to higher computing education, so developments in the field at lower levels of education were not addressed. The manual filtering and searches for the ethical and societal impacts was time consuming, but an effort we believed justified, as limiting the search by terms such as “ethics” or “social impacts” would have missed the bulk of the insights this review was able to glean from the nearly 400 articles looked at in-depth.

We worked through the papers, led by an initial protocol, both individually and as a team, with a data extraction process guided by research questions one and three and an initial set of codes from Martin et al., [227] and those that emerged as the review continued. We consulted regularly to agree and refine our approach to analysis. There will no doubt have been some inconsistencies, but we have extracted the key codes and themes from the data. Future work can extend the scope and analyze this topic with greater depth.

Based on our analysis there was an evidenced excitement and enthusiasm for GenAI technology demonstrated by the volume of literature and illustrated by the top code of 71 papers in Figure 5 showing a positive ‘contribution to learning’. In contrast there was a lesser disciplinary focus on ethical and societal impacts, as critiqued by key figures in the ACM and IEEE professional societies [79, 260, 348] and so we highlight a likely publication bias, towards the reporting of positive results from use of the technology which should be considered.

Our policy evaluation was limited to a set of 21 institutions, leaving room for further research in this area. While we assume that gathering examples of policies with international representation would provide us with a comprehensive view of policies, we acknowledge that this may or may not actually hold true. With GenAI use expanded rapidly in teaching and learning both in and outside the walls of the institution, this policy analysis is but a snapshot at a moment in time. Moreover, with the transition already underway to agentic AI [191], we anticipate that policies should and will shift as time progresses and institutions respond accordingly.

The ESI-Framework draws on multiple social levels and stakeholders positions, however they are not visually represented in Table 6. Instead, these perspectives are activated by the user when applying the framework to a specific dilemma or scenario. On one side the design of the ESI-framework provides flexibility but on the other side it places interpretative responsibility on the user. Hence the effectiveness of applying the framework may depend on the user’s familiarity with computing education, ethical reasoning, stakeholder awareness and contextual understanding.

The ESI-framework is grounded in an extensive literature review, policy evaluation, and expert reflection. However, it has not yet been systematically tested with a wide range of stakeholder groups or across different educational settings.

## 6 Future Work

The ESI-Framework paired with dilemma analysis highlights the need for future research to take critical, interdisciplinary approaches that take into account micro, meso, macro levels [223] and the broader economic, political, social, socio-technical, and historical context [105, 110]. It is also crucial to attend to the crucial connections between and among the macro, meso, and micro levels since attending to these connections provides a lens into the power dynamics at play between and among these levels, providing insights into the root cause of competing priorities. Exploring additional dilemma archetypes and tensions while leveraging the ESI-Framework as a ‘compass’ opens a way to navigate the GenAI landscape with intentionality and deliberation. At the same time, it is crucial to note that the ESI-Framework is intended to be dynamic and responsive with room for refinement in the face of new findings and evidence.

The emerging code tentatively named *Ethical Learning* (Sect. 2.1.5) has been elicited from the set of Educational Insights. As a potential ethical risk directly linked to higher computing education (if not more broadly), *Ethical Learning* points to an area ripe for future investigation.

The current set of dilemma archetypes is not exhaustive. These three ones are intended to demonstrate how the ESI-Framework can be applied to real-life scenarios. Further validation involving stakeholder testing and practitioner narratives and reflections, as well as empirical data drawn from the ever growing body of literature, may help develop more archetypes and contribute to strengthening the reliability and applicability of the framework.

Applying discourse theory to computing education dilemmas and case studies could lead to insights. Documenting stakeholders’ discourse as they navigate tensions and dilemmas in decision-making around computing pedagogy and content aligns with the needs to explore the nuanced ways in which difficult decisions are made. Insights from multidisciplinary research (e.g., human factors, cognitive, technical) [215] that is grounded in the ESI-Framework within the context of dilemma analysis could lead to a deeper understanding of the intersecting factors that define users’ experience of AI systems. This type of analysis may yield important insights into how stakeholders weigh benefits vs. risks, opportunities vs. challenges, and harm vs. good.

# 7 Conclusion

In this paper, we have investigated the impacts of GenAI on higher computing education and outlined the results of a systematic literature review (addressing RQ1), an evaluation of a broad set of international policies that illuminates the socio-technical dynamics operating in higher education institutions (addressing RQ2), and the development and validation of the ESI-Framework through dilemma analysis (addressing RQ3). Collectively, these results aim to inform research, practice and policy development in this rapidly evolving field. The ESI-Framework developed through this project is intended to guide computing education faculty and institutions navigate this complex landscape, providing a framework that acknowledges the tensions and nuances behind both creating and using GenAI. The ACM Code of Ethics and Professional Conduct begins by stating that "Computing professionals' actions change the world" [119]. The need to build frameworks that support deliberation and reflection on the ethical and societal impacts of computing to support the public good is urgent and pressing, with technologies impacting access to and experience of education, employment, healthcare, and voting [2]. Indeed, the principles within the ACM Code are presented as statements of responsibility based on the premise that designing and creating technologies for the public good is of primary importance. The ESI-Framework and dilemma analysis provide a framework for the higher education computing community to engage in open discussions about ethical issues to advance fairness, accountability, transparency, inclusion, and trust through a human-centered, decision-making process toward human flourishing for all.

# Acknowledgments

This work is funded through the Association for Computing Machinery (ACM) Education Advisory Committee.

**Table 11: List of references in the initial coded set for literature review.**

| First Author_Year | Title | Database | Citation |
|---|---|---|---|
| Budhiraja_2024 | It's not like Jarvis, but it's pretty close! - Examining ChatGPT's Usage among Undergraduate Students in Computer Science | ACM | [51] |
| Prather_2023 | It's Weird That it Knows What I Want: Usability and Interactions with Copilot for Novice Programmers | ACM | [278] |
| Bernstein_2024 | "Like a Nesting Doll": Analyzing Recursion Analogies Generated by CS Students Using Large Language Models | ACM | [39] |
| Kasinidou_2024 | "We have to learn to work with such systems": Students' Perceptions of ChatGPT After a Short Educational Intervention on NLP | ACM | [175] |
| Zamfirescu-Pereira_2025 | 61A Bot Report: AI Assistants in CS1 Save Students Homework Time and Reduce Demands on Staff. (Now What?) | ACM | [373] |
| Farinetti_2025 | A Critical Approach to ChatGPT: An Experience in SQL Learning | ACM | [96] |
| Moreira_2025 | A Roadmap for Integrating Sustainability into Software Engineering Education | ACM | [244] |
| Vadaparty_2025 | Achievement Goals in CS1-LLM | ACM | [345] |
| Vahid_2024 | AI in CS Education: Opportunities, Challenges, and Pitfalls to Avoid | ACM | [346] |
| Woodrow_2024 | AI Teaches the Art of Elegant Coding: Timely, Fair, and Helpful Style Feedback in a Global Course | ACM | [361] |
| Kendon_2023 | AI-Generated Code Not Considered Harmful | ACM | [180] |
| Minagar_2024 | ALAN: Assessment-as-Learning Authentic Tasks for Networking | ACM | [239] |
| Mezzaro_2024 | An Empirical Study on How Large Language Models Impact Software Testing Learning | ACM | [235] |
| Rasnayaka_2024 | An Empirical Study on Usage and Perceptions of LLMs in a Software Engineering Project | ACM | [292] |
| Feng_2024 | An Eye for an AI: Evaluating GPT-4o's Visual Perception Skills and Geometric Reasoning Skills Using Computer Graphics Questions | ACM | [98] |
| Arora_2025 | Analyzing LLM Usage in an Advanced Computing Class in India | ACM | [22] |
| Basit_2025 | ASCI: AI-Smart Classroom Initiative | ACM | [34] |
| Ohm_2024 | Assessing the Impact of Large Language Models on Cybersecurity Education: A Study of ChatGPT's Influence on Student Performance | ACM | [253] |
| Rogers_2024 | Attitudes Towards the Use (and Misuse) of ChatGPT: A Preliminary Study | ACM | [299] |
| Boubaker_2024 | Automated Generation of Challenge Questions for Student Code Evaluation Using Abstract Syntax Tree Embeddings and RAG: An Exploratory Study | ACM | [47] |
| Richards_2024 | Bob or Bot: Exploring ChatGPT's Answers to University Computer Science Assessment | ACM | [296] |
| Prather_2025 | Breaking the Programming Language Barrier: Multilingual Prompting to Empower Non-Native English Learners | ACM | [281] |
| Yeh_2025 | Bridging Novice Programmers and LLMs with Interactivity | ACM | [370] |
| Padurean_2025 | BugSpotter: Automated Generation of Code Debugging Exercises | ACM | [272] |
| Golesteanu_2024 | Can ChatGPT pass a Theory of Computing Course? | ACM | [116] |
| Al-Hossami_2024 | Can Language Models Employ the Socratic Method? Experiments with Code Debugging | ACM | [9] |
| Pang_2024 | ChatGPT and Cheat Detection in CS1 Using a Program Autograding System | ACM | [268] |
| Zheng_2023 | ChatGPT for Teaching and Learning: An Experience from Data Science Education | ACM | [377] |
| Qureshi_2023 | ChatGPT in Computer Science Curriculum Assessment: An analysis of Its Successes and Shortcomings | ACM | [285] |
| Joshi_2024 | ChatGPT in the Classroom: An Analysis of Its Strengths and Weaknesses for Solving Undergraduate Computer Science Questions | ACM | [166] |
| Ouh_2023 | ChatGPT, Can You Generate Solutions for my Coding Exercises? An Evaluation on its Effectiveness in an undergraduate Java Programming Course. | ACM | [258] |
| Kazemitabaar_2024 | CodeAid: Evaluating a Classroom Deployment of an LLM-based Programming Assistant that Balances Student and Educator Needs | ACM | [178] |

| First Author_Year | Title | Database | Citation |
|---|---|---|---|
| Leinonen_2023 | Comparing Code Explanations Created by Students and Large Language Models | ACM | [203] |
| Nguyen_2024 | Comparing Feedback from Large Language Models and Instructors: Teaching Computer Science at Scale | ACM | [251] |
| Renzella_2025 | Compiler-Integrated, Conversational AI for Debugging CS1 Programs | ACM | [295] |
| Denny_2024 | Computing Education in the Era of Generative AI | ACM | [82] |
| Fernandez_2024 | CS1 with a Side of AI: Teaching Software Verification for Secure Code in the Era of Generative AI | ACM | [102] |
| Taylor_2024 | dcc –help: Transforming the Role of the Compiler by Generating Context-Aware Error Explanations with Large Language Models | ACM | [335] |
| Yang_2024 | Debugging with an AI Tutor: Investigating Novice Help-seeking Behaviors and Perceived Learning | ACM | [369] |
| Buffardi_2024 | Designing a CURE for CS1 | ACM | [52] |
| Hoq_2024 | Detecting ChatGPT-Generated Code Submissions in a CS1 Course Using Machine Learning Models | ACM | [143] |
| Sakzad_2024 | Diverging assessments: What, Why, and Experiences | ACM | [305] |
| Padiyath_2024 | Do I Have a Say in This, or Has ChatGPT Already Decided for Me? | ACM | [262] |
| Smith_2024 | Early Adoption of Generative Artificial Intelligence in Computing Education: Emergent Student Use Cases and Perspectives in 2023 | ACM | [324] |
| Pang_2024 | Examining the Relationship between Socioeconomic Status and Beliefs about Large Language Models in an Undergraduate Programming Course | ACM | [268] |
| MacNeil_2023 | Experiences from Using Code Explanations Generated by Large Language Models in a Web Software Development E-Book | ACM | [219] |
| Denny_2024 | Explaining Code with a Purpose: An Integrated Approach for Developing Code Comprehension and Prompting Skills | ACM | [82] |
| Rajabi_2023 | Exploring ChatGPTÕs impact on post-secondary education: A qualitative study | ACM | [289] |
| Fenu_2024 | Exploring Student Interactions with AI in Programming Training | ACM | [100] |
| Kerslake_2025 | Exploring Student Reactions to LLM-Generated Feedback on Explain in Plain English Problems | ACM | [182] |
| Schefer-Wenzl_2024 | Exploring the Adoption of Generative AI Tools in Computer Science Education: A Student Survey | ACM | [307] |
| Kazemitabaar_2025 | Exploring the Design Space of Cognitive Engagement Techniques with AI-Generated Code for Enhanced Learning | ACM | [177] |
| Hellas_2023 | Exploring the Responses of Large Language Models to Beginner ProgrammersÕ Help Requests | ACM | [139] |
| Ahmed_2025 | Feasibility Study of Augmenting Teaching Assistants with AI for CS1 Programming Feedback | ACM | [5] |
| Azaiz_2024 | Feedback-Generation for Programming Exercises With GPT-4 | ACM | [26] |
| Apiola_2024 | First Year CS Students Exploring And Identifying Biases and Social Injustices in Text-to-Image Generative AI | ACM | [21] |
| Biderman_2022 | Fooling MOSS Detection with Pretrained Language Models | ACM | [41] |
| Filcik_2025 | Fostering Creativity: Student-Generative AI Teaming in an Open-Ended CS0 Assignment | ACM | [103] |
| Lau_2023 | From "Ban It Till We Understand It" to "Resistance is Futile": How University Programming Instructors Plan to Adapt as More Students Use AI Code Generation and Explanation Tools such as ChatGPT and GitHub Copilot | ACM | [198] |
| Feng_2025 | From Automation to Cognition: Redefining the Roles of Educators and Generative AI in Computing Education | ACM | [99] |
| Prasad_2023 | Generating Programs Trivially: Student Use of Large Language Models | ACM | [277] |
| Villegas_2024 | Generation and Evaluation of a Culturally-Relevant CS1 Textbook for Latines using Large Language Models | ACM | [349] |
| Poitras_2024 | Generative AI in Introductory Programming Instruction: Examining the Assistance Dilemma with LLM-Based Code Generators | ACM | [274] |

| First Author_Year | Title | Database | Citation |
|---|---|---|---|
| Mahon_2024 | Guidelines for the Evolving Role of Generative AI in Introductory Programming Based on Emerging Practice | ACM | [221] |
| Kumar_2024 | Guiding Students in Using LLMs in Supported Learning Environments: Effects on Interaction Dynamics, Learner Performance, Confidence, and Trust | ACM | [193] |
| Nguyen_2024 | How Beginning Programmers and Code LLMs (Mis)read Each Other | ACM | [251] |
| Daun_2023 | How ChatGPT Will Change Software Engineering Education | ACM | [77] |
| Shen_2024 | Implications of ChatGPT for Data Science Education | ACM | [318] |
| Shochcho_2025 | Improving User Engagement and Learning Outcomes in LLM-Based Python Tutor: A Study of PACE | ACM | [320] |
| Padiyath_2024 | Insights from Social Shaping Theory: The Appropriation of Large Language Models in an Undergraduate Programming Course | ACM | [262] |
| Sheard_2024 | Instructor Perceptions of AI Code Generation Tools - A Multi-Institutional Interview Study | ACM | [316] |
| Ma_2024 | Integrating AI Tutors in a Programming Course | ACM | [214] |
| Oran_2024 | Integrating ChatGPT in Project Management Education: Benefits and Challenges in the Academic Environment | ACM | [255] |
| Kerslake_2024 | Integrating Natural Language Prompting Tasks in Introductory Programming Courses | ACM | [183] |
| Yu_2025 | Integrating Small Language Models with Retrieval-Augmented Generation in Computing Education: Key Takeaways, Setup, and Practical Insights | ACM | [371] |
| Li_2025 | Investigating the Capabilities of Generative AI in Solving Data Structures, Algorithms, and Computability Problems | ACM | [205] |
| Balse_2023 | Investigating the Potential of GPT-3 in Providing Feedback for Programming Assessments | ACM | [31] |
| Dunder_2024 | Kattis vs ChatGPT: Assessment and Evaluation of Programming Tasks in the Age of Artificial Intelligence | ACM | [91] |
| Clift_2025 | Learning without Limits: Analysing the Usage of Generative AI in a Summative Assessment | ACM | [70] |
| Lehtinen_2024 | Let's Ask AI About Their Programs: Exploring ChatGPT's Answers To Program Comprehension Questions | ACM | [200] |
| Hyrynsalmi_2025 | Making Software Development More Diverse and Inclusive: Key Themes, Challenges, and Future Directions | ACM | [151] |
| Haji_2025 | Midterm Exam Outliers Efficiently Highlight Potential Cheaters on Programming Assignments | ACM | [129] |
| Feng_2024 | More Than Meets the AI: Evaluating the performance of GPT-4 on Computer Graphics assessment questions | ACM | [98] |
| Jordan_2024 | Need a Programming Exercise Generated in Your Native Language? ChatGPT's Got Your Back: Automatic Generation of Non-English Programming Exercises Using OpenAI GPT-3.5 | ACM | [165] |
| Santos_2024 | Not the Silver Bullet: LLM-enhanced Programming Error Messages are Ineffective in Practice | ACM | [306] |
| AlOmar_2025 | Nurturing Code Quality: Leveraging Static Analysis and Large Language Models for Software Quality in Education | ACM | [14] |
| Malinka_2023 | On the Educational Impact of ChatGPT: Is Artificial Intelligence Ready to Obtain a University Degree? | ACM | [224] |
| Russell_2023 | Online Programming Exams - An Experience Report | ACM | [301] |
| Qiao_2025 | Oversight in Action: Experiences with Instructor-Moderated LLM Responses in an Online Discussion Forum | ACM | [284] |
| Gardella_2024 | Performance, Workload, Emotion, and Self-Efficacy of Novice Programmers Using AI Code Generation | ACM | [109] |
| Chen_2024 | Plagiarism in the Age of Generative AI: Cheating Method Change and Learning Loss in an Intro to CS Course | ACM | [62] |

| First Author_Year | Title | Database | Citation |
|---|---|---|---|
| Becker_2023 | Programming Is Hard - Or at Least It Used to Be: Educational Opportunities and Challenges of AI Code Generation | ACM | [35] |
| Smith_2024 | Prompting for Comprehension: Exploring the Intersection of Explain in Plain English Questions and Prompt Writing | ACM | [324] |
| Koutcheme_2024 | Propagating Large Language Models Programming Feedback | ACM | [192] |
| Riello_2024 | Reimagining Student Success Prediction: Applying LLMs in Educational AI with XAI | ACM | [297] |
| Hazzan_2025 | Rethinking Computer Science Education in the Age of GenAI | ACM | [136] |
| Cao_2023 | Scaffolding CS1 Courses with a Large Language Model-Powered Intelligent Tutoring System | ACM | [56] |
| Gutierrez_2025 | Seeing the Forest and the Trees: Solving Visual Graph and Tree Based Data Structure Problems using Large Multimodal Models | ACM | [126] |
| Margulieux_2024 | Self-Regulation, Self-Efficacy, and Fear of Failure Interactions with How Novices Use LLMs to Solve Programming Problems | ACM | [226] |
| Nelson_2025 | SENSAI: Large Language Models as Applied Cybersecurity Tutors | ACM | [249] |
| Jegourel_2024 | Sieving Coding Assignments Over Submissions Generated by AI and Novice Programmers | ACM | [161] |
| Tang_2025 | SPHERE: Supporting Personalized Feedback at Scale in Programming Classrooms with Structured Review of Generative AI Outputs | ACM | [333] |
| Amoozadeh_2024 | Student-AI Interaction: A Case Study of CS1 students | ACM | [16] |
| Keuning_2024 | Students' Perceptions and Use of Generative AI Tools for Programming Across Different Computing Courses | ACM | [184] |
| Shah_2025 | Students' Use of GitHub Copilot for Working with Large Code Bases | ACM | [314] |
| Denzler_2024 | Style Anomalies Can Suggest Cheating in CS1 Programs | ACM | [83] |
| MacNeil_2024 | Synthetic Students: A Comparative Study of Bug Distribution Between Large Language Models and Computing Students | ACM | [217] |
| Dingle_2024 | Tackling Students' Coding Assignments with LLMs | ACM | [85] |
| Jacques_2023 | Teaching CS-101 at the Dawn of ChatGPT | ACM | [156] |
| Gumina_2023 | Teaching IT Software Fundamentals: Strategies and Techniques for Inclusion of Large Language Models: Strategies and Techniques for Inclusion of Large Language Models | ACM | [124] |
| Bouvier_2024 | Teaching Programming Error Message Understanding | ACM | [48] |
| Bopp_2024 | The Case for LLM Workshops | ACM | [46] |
| Hou_2024 | The Effects of Generative AI on Computing StudentsÕ Help-Seeking Preferences | ACM | [146] |
| Prather_2023 | The Robots Are Here: Navigating the Generative AI Revolution in Computing Education | ACM | [278] |
| McGowan_2024 | The use of ChatGPT to generate Summative Feedback in Programming Assessments that is Consistent, Prompt, without Bias and Scalable | ACM | [230] |
| Prather_2024 | The Widening Gap: The Benefits and Harms of Generative AI for Novice Programmers | ACM | [282] |
| Moudgalya_2024 | Toward Data Sovereignty: Justice-oriented and Community-based AI Education | ACM | [245] |
| Xiang_2023 | Toward Reproducing Network Research Results Using Large Language Models | ACM | [362] |
| Vahid_2024 | Towards Comprehensive Metrics for Programming Cheat Detection | ACM | [346] |
| Amoozadeh_2024 | Trust in Generative AI among Students: An exploratory study | ACM | [16] |
| Ramirez_2025 | Understanding the Impact of Using Generative AI Tools in a Database Course | ACM | [290] |
| Servin_2024 | Unfolding Programming: How to Use AI Tools in Introductory Computing Courses | ACM | [312] |
| Gorson_2025 | Unlocking Potential with Generative AI Instruction: Investigating Mid-level Software Development Student Perceptions, Behavior, and Adoption | ACM | [118] |
| Korpimies_2024 | Unrestricted Use of LLMs in a Software Project Course: Student Perceptions on Learning and Impact on Course Performance | ACM | [190] |
| Rönnberg_2024 | Use of Generative AI for Fictional Field Studies in Design Courses | ACM | [300] |
| Nguyen_2024 | Using GPT-4 to Provide Tiered, Formative Code Feedback | ACM | [251] |

| First Author_Year | Title | Database | Citation |
|---|---|---|---|
| Leinonen_2023 | Using Large Language Models to Enhance Programming Error Messages | ACM | [203] |
| Jamie_2025 | Utilizing ChatGPT in a Data Structures and Algorithms Course: A Teaching Assistant's Perspective | ACM | [157] |
| Thorgeirsson_2025 | What Can Computer Science Educators Learn From the Failures of Top-Down Pedagogy? | ACM | [337] |
| Kiesler_2024 | With Great Power Comes Great Responsibility - Integrating Data Ethics into Computing Education | ACM | [186] |
| Neumann_2025 | An LLM-Driven Chatbot in Higher Education for Databases and Information Systems | IEEE | [250] |
| Jo_2023 | Analysis of Plagiarism via ChatGPT on Domain-Specific Exams | IEEE | [164] |
| Gerede_2024 | Are We Asking the Right Questions to ChatGPT for Learning Software Design Patterns? | IEEE | [111] |
| Hajj_2023 | Assessing the Impact of ChatGPT in a PHP Programming Course | IEEE | [130] |
| Murali_2024 | Augmenting Virtual Labs with Artificial Intelligence for Hybrid Learning | IEEE | [247] |
| SaÇ§lam_2024 | Automated Detection of AI-Obfuscated Plagiarism in Modeling Assignments | IEEE | [303] |
| Vintila_2024 | AVERT (Authorship Verification and Evaluation Through Responsive Testing): an LLM-Based Procedure that Interactively Verifies Code Authorship and Evaluates Student Understanding | IEEE | [350] |
| Weber_2024 | Beyond the Hype: Perceptions and Realities of Using Large Language Models in Computer Science Education at an R1 University | IEEE | [355] |
| Brach_2024 | Can Large Language Model Detect Plagiarism in Source Code? | IEEE | [49] |
| Dos_2023 | Challenging the Confirmation Bias: Using ChatGPT as a Virtual Peer for Peer Instruction in Computer Programming Education | IEEE | [87] |
| Barambones_2024 | ChatGPT for Learning HCI Techniques: A Case Study on Interviews for Personas | IEEE | [33] |
| Nutalapati_2024 | Coding Buddy: An Adaptive AI-Powered Platform for Personalized Learning | IEEE | [252] |
| Weichert_2024 | Computer Science Student Attitudes Towards AI Ethics and Policy: A Preliminary Investigation | IEEE | [356] |
| Orenstrakh_2024 | Detecting LLM-Generated Text in Computing Education: Comparative Study for ChatGPT Cases | IEEE | [256] |
| Kalluri_2024 | Developing Future Computational Thinking in Foundational CS Education: A Case Study From a Liberal Education University in India | IEEE | [168] |
| Xue_2024 | Does ChatGPT Help With Introductory Programming? An Experiment of Students Using ChatGPT in CS1 | IEEE | [366] |
| Hean_2024 | Enhancing the Teaching of Data Structures and Algorithms using AI Chatbots | IEEE | [137] |
| Andonov_2024 | Enhancing University Students' Activities with Interactive Immediate Feedback Using Customized LLMs | IEEE | [19] |
| Jacobs_2024 | Evaluating the Application of Large Language Models to Generate Feedback in Programming Education | IEEE | [155] |
| Fwa_2024 | Experience Report: Identifying Common Misconceptions and Errors of Novice Programmers with ChatGPT | IEEE | [107] |
| Palacios-Alonso_2024 | Experiences and Proposals of Use of Generative AI in Advanced Software Courses | IEEE | [265] |
| Hu_2023 | Explicitly Introducing ChatGPT into First-year Programming Practice: Challenges and Impact | IEEE | [147] |
| Hingle_2023 | Exploring NLP-Based Methods for Generating Engineering Ethics Assessment Qualitative Codebooks | IEEE | [141] |
| Kiesler_2023 | Exploring the Potential of Large Language Models to Generate Formative Programming Feedback | IEEE | [185] |
| Rafiee_2024 | Fostering Personalized Learning in Data Science: Integrating Innovative Tools and Strategies for Diverse Pathways | IEEE | [287] |
| Zastudil_2023 | Generative AI in Computing Education: Perspectives of Students and Instructors | IEEE | [374] |
| CÃ¡mara_2024 | Generative AI in the Software Modeling Classroom: An Experience Report With ChatGPT and Unified Modeling Language | IEEE | [53] |

| First Author_Year | Title | Database | Citation |
|---|---|---|---|
| Cubillos_2025 | Generative Artificial Intelligence in Computer Programming: Does It Enhance Learning, Motivation, and the Learning Environment? | IEEE | [72] |
| Choudhuri_2024 | How Far Are We? The Triumphs and Trials of Generative AI in Learning Software Engineering | IEEE | [65] |
| Scholl_2024 | How Novice Programmers Use and Experience ChatGPT when Solving Programming Exercises in an Introductory Course | IEEE | [308] |
| Akbar_2024 | Human Evaluation of GPT for Scalable Python Programming Exercise Generation | IEEE | [7] |
| Tracy_2025 | Impact of GPT-Driven Teaching Assistants in VR Learning Environments | IEEE | [341] |
| Shaer_2024 | Integrating Generative Artificial Intelligence to a Project-Based Tangible Interaction Course | IEEE | [313] |
| Tayeb_2024 | Investigating Developers' Preferences for Learning and Issue Resolution Resources in the ChatGPT Era | IEEE | [334] |
| Zhang_2024 | JavaLLM: A Fine-Tuned LLM for Java Programming Education | IEEE | [376] |
| Jacobs_2024 | Leveraging Lecture Content for Improved Feedback: Explorations with GPT-4 and Retrieval Augmented Generation | IEEE | [155] |
| Liu_2023 | Leveraging the Power of AI in Undergraduate Computer Science Education: Opportunities and Challenges | IEEE | [209] |
| Rachha_2024 | LLM-Enhanced Learning Environments for CS: Exploring Data Structures and Algorithms with Gurukul | IEEE | [286] |
| Hang_2024 | MCQGen: A Large Language Model-Driven MCQ Generator for Personalized Learning | IEEE | [133] |
| Kumar_2024 | Optimizing Large Language Models for Auto-Generation of Programming Quizzes | IEEE | [193] |
| Simaremare_2024 | Pair Programming in Programming Courses in the Era of Generative AI: Students' Perspective | IEEE | [322] |
| Karnalim_2023 | Plagiarism and AI Assistance Misuse in Web Programming: Unfair Benefits and Characteristics | IEEE | [174] |
| Kumar_2024 | Preliminary Results from Integrating Chatbots and Low-Code AI in Computer Science Coursework | IEEE | [193] |
| Berrezueta-Guzman_2023 | Recommendations to Create Programming Exercises to Overcome ChatGPT | IEEE | [40] |
| Wang_2023 | Student Mastery or AI Deception? Analyzing ChatGPT's Assessment Proficiency and Evaluating Detection Strategies | IEEE | [354] |
| Takerngsaksiri_2024 | Students' Perspectives on AI Code Completion: Benefits and Challenges | IEEE | [331] |
| Moore_2024 | Sustainability, Ethics and Artificial Intelligence in Computing Education | IEEE | [242] |
| Cao_2024 | The Impact of Role Strategy on ChatGPT-Assisted Asynchronous Online Discussions: An Experimental Study | IEEE | [57] |
| Banerjee_2025 | Understanding ChatGPT: Impact Analysis and Path Forward for Teaching Computer Science and Engineering | IEEE | [32] |
| Shin_2024 | Understanding Optimal Interactions Between Students and A Chatbot During A Programming Task | IEEE | [319] |
| Andersen-Kiel_2024 | Using ChatGPT in Undergraduate Computer Science and Software Engineering Courses: A Students' Perspective | IEEE | [18] |
| Randall_2024 | What an AI-Embracing Software Engineering Curriculum Should Look Like: An Empirical Study | IEEE | [291] |
| Crandall_2024 | WIP: ARTful Insights from a Pilot Study on GPT-Based Automatic Code Reviews in Undergraduate Computer Science Programs | IEEE | [71] |
| Haldar_2024 | WIP: Assessing the Effectiveness of ChatGPT in Preparatory Testing Activities | IEEE | [132] |
| Ilic_2024 | WIP: Investigating the Use of AI Chatbots by Undergraduate Computer Science Students | IEEE | [152] |
| Jamieson_2023 | With ChatGPT, Do We have to Rewrite Our Learning Objectives - CASE Study in Cybersecurity | IEEE | [158] |

| First Author_Year | Title | Database | Citation |
|---|---|---|---|
| Lauren_2023 | Work-in-Progress: Integrating Generative AI with Evidence-based Learning Strategies in Computer Science and Engineering Education | IEEE | [199] |
| Mailach_2025 | Ok Pal, we have to code that now: interaction patterns of programming beginners with a conversational chatbot | scopus | [222] |
| Doyle_2025 | A comparative study of AI-generated and human-crafted learning objectives in computing education | scopus | [89] |
| Maurat_2024 | A Comparative Study of Gender Differences in the Utilization and Effectiveness of AI-Assisted Learning Tools in Programming Among University Students | scopus | [228] |
| Oyelere_2025 | A comparative study of student perceptions on generative AI in programming education across Sub-Saharan Africa | scopus | [259] |
| Vainionp_2024 | A nexus analysis of future ICT professionals' views on sustainable digital technology development | scopus | [347] |
| Michelutti_2024 | A Systematic Study on the Potentials and Limitations of LLM-assisted Software Development | scopus | [237] |
| Azoulay_2024 | Academia and Industry Synergy: Addressing Integrity Challenge in Programming Education | scopus | [27] |
| Lopez-Fernndez_2024 | Adoption and Impact of ChatGPT in Computer Science Education: A Case Study on a Database Administration Course | scopus | [211] |
| Akapõnar_2024 | AI chatbots in programming education: guiding success or encouraging plagiarism | scopus | [8] |
| Gottipati_2023 | AI for Connectivism Learning - Undergraduate Students' Experiences of ChatGPT in Advanced Programming Courses | scopus | [121] |
| Tang_2024 | AI-Generated Programming Solutions: Impacts on Academic Integrity and Good Practices | scopus | [332] |
| Dickey_2024 | AI-Lab: A Framework for Introducing Generative Artificial Intelligence Tools in Computer Programming Courses | scopus | [84] |
| Frankford_2024 | AI-Tutoring in Software Engineering Education Experiences with Large Language Models in Programming Assessments | scopus | [106] |
| Jamil_2025 | An Experiment with LLMs as Database Design Tutors: Persistent Equity and Fairness Challenges in Online Learning | scopus | [159] |
| Lai_2025 | Analysis of Learning Behaviors and Outcomes for Students with Different Knowledge Levels: A Case Study of Intelligent Tutoring System for Coding and Learning (ITS-CAL) | scopus | [196] |
| Hasan_2024 | Analyzing the uses and perceptions of computer science students towards generative AI tools | scopus | [135] |
| Habiballa_2025 | Artificial Intelligence (ChatGPT) and Bloom's Taxonomy in Theoretical Computer Science Education | scopus | [127] |
| Eaton_2024 | Artificial Intelligence in the CS2023 Undergraduate Computer Science Curriculum: Rationale and Challenges | scopus | [93] |
| Hartley_2024 | Artificial Intelligence Supporting Independent Student Learning: An Evaluative Case Study of ChatGPT and Learning to Code | scopus | [134] |
| Pan_2024 | Assessing AI Detectors in Identifying AI-Generated Code: Implications for Education | scopus | [267] |
| Troussas_2024 | Assessing the Impact of Integrating ChatGPT as an Advice Generator in Educational Software | scopus | [343] |
| Dai_2024 | Assessing the proficiency of large language models in automatic feedback generation: An evaluation study | scopus | [74] |
| Alkafaween_2025 | Automating Autograding: Large Language Models as Test Suite Generators for Introductory Programming | scopus | [13] |
| Singh_2024 | Bridging Learnersourcing and AI: Exploring the Dynamics of Student-AI Collaborative Feedback Generation | scopus | [323] |
| Humphrys_2025 | Bringing AI APIs into the Classroom with a JavaScript Coding Site | scopus | [150] |

| First Author_Year | Title | Database | Citation |
|---|---|---|---|
| Elentukh_2025 | Building a Chatbot to Adopt an Effective Learning Strategy for Graduate Courses in Computer Science | scopus | [94] |
| Kofahi_2025 | CHATGPT FOR OPERATING SYSTEMS: HIGHER-ORDER THINKING IN FOCUS | scopus | [6] |
| Lpez-Fernndez_2025 | ChatGPT in Computer Science Education: A Case Study on a Database Administration Course | scopus | [212] |
| Kim_2024 | ChatGPT in Data Visualization Education: A Student Perspective | scopus | [187] |
| Popovici_2024 | ChatGPT in the Classroom. Exploring Its Potential and Limitations in a Functional Programming Course | scopus | [275] |
| Humble_2024 | Cheaters or AI-Enhanced Learners: Consequences of ChatGPT for Programming Education | scopus | [149] |
| Rebollido_2024 | Code Comprehension Problems inÊIntroductory Programming toÊOvercome ChatGPT | scopus | [294] |
| Torres_2025 | CodeContrast: A Contrastive Learning Approach for Generating Coherent Programming Exercises | scopus | [340] |
| Ghimire_2024 | Coding withÊAI: How Are Tools Like ChatGPT Being Used by Students in Foundational Programming Courses | scopus | [113] |
| Kanabar_2025 | Cross-Continental Insights: Comparative Analysis of Using AI for Information System Stakeholder Analysis in Undergraduate Courses in the EU and USA | scopus | [169] |
| Agrawal_2024 | CyberQ: Generating Questions and Answers for Cybersecurity Education Using Knowledge Graph-Augmented LLMs | scopus | [4] |
| Grtl_2025 | Design and Evaluation of an LLM-Based Mentor for Software Architecture in Higher Education Project Management Classes | scopus | [125] |
| Haindl_2024 | Does ChatGPT Help Novice Programmers Write Better Code? Results From Static Code Analysis | scopus | [128] |
| Lepp_2025 | Does generative AI help in learning programming: Students' perceptions, reported use and relation to performance | scopus | [204] |
| Shanshan_2024 | Empowering learners with AI-generated content for programming learning and computational thinking: The lens of extended effective use theory | scopus | [315] |
| Soto_2024 | Enhancing Learning Dynamics: Integrating Interactive Learning Environments and ChatGPT for Computer Networking Lessons | scopus | [325] |
| Ma_2024 | Enhancing Programming Education withÊChatGPT: A Case Study onÊStudent Perceptions andÊInteractions inÊaÊPython Course | scopus | [214] |
| Yang_2024 | Enhancing python learning with PyTutor: Efficacy of a ChatGPT-Based intelligent tutoring system in programming education | scopus | [369] |
| Mendonça_2025 | Evaluating LLMs for Automated Scoring in Formative Assessments | scopus | [233] |
| Dannath_2024 | Evaluating Task-Level Struggle Detection Methods in Intelligent Tutoring Systems for Programming | scopus | [76] |
| Dirin_2024 | Examining the Utilization of Artificial Intelligence Tools by Students in Software Engineering Projects | scopus | [86] |
| Xiao_2024 | Exploring How Multiple Levels of GPT-Generated Programming Hints Support or Disappoint Novices | scopus | [364] |
| Ma_2024 | EXPLORING STUDENT PERCEPTION AND INTERACTION USING CHATGPT IN PROGRAMMING EDUCATION | scopus | [214] |
| Mellqvist_2024 | Exploring Student Perspectives on Generative AI in Requirements Engineering Education | scopus | [232] |
| Gottipati_2023 | Exploring Students' Adoption of ChatGPT as a Mentor for Undergraduate Computing Projects: PLS-SEM Analysis | scopus | [121] |
| Halani_2025 | Exploring the Effectiveness of a Multilingual Generative AI Programming Chatbot for Vernacular Medium CS Students | scopus | [131] |
| Fernandez-y-Fernandez_2024 | Exploring the Frontier of Software Engineering Education with Chatbots | scopus | [367] |
| Troussas_2025 | Fuzzy Memory Networks and Contextual Schemas: Enhancing ChatGPT Responses in a Personalized Educational System | scopus | [342] |

| First Author_Year | Title | Database | Citation |
|---|---|---|---|
| Kelly_2024 | Gamification Powered by a Large Language Model to Enhance Flipped Classroom Learning in Undergraduate Computer Science | scopus | [179] |
| Kumar_2024 | GenAI Tools to Improve Data Science Project Outcomes | scopus | [193] |
| Raybourn_2023 | Guidelines for Practicing Responsible Innovation in HPC: A Sociotechnical Approach | scopus | [293] |
| Mohamed_2025 | Hands-on analysis of using large language models for the auto evaluation of programming assignments | scopus | [240] |
| Sun_2025 | Harnessing code domain insights: Enhancing programming Knowledge Tracing with Large Language Models | scopus | [329] |
| Camargo_2025 | Humanities and AI: Ethical Education in Technology Careers | scopus | [54] |
| Toba_2024 | Inappropriate Benefits and Identification of ChatGPT Misuse in Programming Tests: A Controlled Experiment | scopus | [338] |
| Paiva_2025 | Incremental Repair Feedback on Automated Assessment of Programming Assignments | scopus | [263] |
| Palacios_2025 | Influence of ChatGPT on Programming Code Generation: A Case Study of the Technical University of Manabi | scopus | [264] |
| Kargupta_2024 | Instruct, Not Assist: LLM-based Multi-Turn Planning and Hierarchical Questioning for Socratic Code Debugging | scopus | [170] |
| Katona_2025 | Integrating AI-based adaptive learning into the flipped classroom model to enhance engagement and learning outcomes | scopus | [176] |
| Sajja_2024 | Integrating Generative AI in Hackathons: Opportunities, Challenges, and Educational Implications | scopus | [304] |
| Bikanga_2024 | It Helps with Crap Lecturers and Their Low Effort: Investigating Computer Science StudentsÕ Perceptions of Using ChatGPT for Learning | scopus | [43] |
| Budhiraja_2024 | It's not like Jarvis, but it's pretty close! - Examining ChatGPT's Usage among Undergraduate Students in Computer Science | scopus | [51] |
| Fehnker_2024 | Keeping Humans inÊtheÊLoop: LLM Supported Oral Examinations | scopus | [97] |
| Leidner_2024 | Language-Model Assisted Learning How toÊProgram? | scopus | [201] |
| Pozdniakov_2024 | Large language models meet user interfaces: The case of provisioning feedback | scopus | [276] |
| Zafar_2025 | Large Language Models: An Empirical Study inÊComputer Science | scopus | [372] |
| Boateng_2024 | Leveraging AI to Advance Science andÊComputing Education Across Africa: Challenges, Progress and Opportunities | scopus | [44] |
| Lohr_2025 | Leveraging Large Language Models to Generate Course-Specific Semantically Annotated Learning Objects | scopus | [210] |
| Sychev_2025 | Mass Generation of Programming Learning Problems from Public Code Repositories | scopus | [330] |
| Gold_2025 | On theÊHelpfulness ofÊaÊZero-Shot Socratic Tutor | scopus | [115] |
| Duong_2024 | ProgEdu4Web: An automated assessment tool for motivating the learning of web programming course | scopus | [92] |
| Boguslawski_2025 | Programming education and learner motivation in the age of generative AI: student and educator perspectives | scopus | [45] |
| Palahan_2025 | PythonPal: Enhancing Online Programming Education Through Chatbot-Driven Personalized Feedback | scopus | [266] |
| Karnalim_2025 | Similarities of Human and AI Assistance in Programming Plagiarism: Student Perspective | scopus | [173] |
| Karnalim_2024 | Simulating Similarities to Maintain Academic Integrity in Programming | scopus | [172] |
| Manikani_2025 | SQL Autograder: Web-based LLM-powered Autograder for Assessment of SQL Queries | scopus | [225] |
| Haindl_2024 | Students' Experiences of Using ChatGPT in an Undergraduate Programming Course | scopus | [128] |
| Zhang_2024 | Students' Perceptions and Preferences of Generative Artificial Intelligence Feedback for Programming | scopus | [376] |

| First Author_Year | Title | Database | Citation |
|---|---|---|---|
| Chen_2025 | StuGPTViz: A Visual Analytics Approach to Understand Student-ChatGPT Interactions | scopus | [63] |
| Jin_2024 | Teach AI How to Code: Using Large Language Models as Teachable Agents for Programming Education | scopus | [162] |
| Menolli_2024 | Teaching Refactoring to Improve Code Quality with ChatGPT: An Experience Report in Undergraduate Lessons | scopus | [234] |
| Aggrawal_2024 | Teamwork Conflict Management Training and Conflict Resolution Practice via Large Language Models | scopus | [3] |
| Jo_t_2024 | The Impact of Large Language Models on Programming Education and Student Learning Outcomes | scopus | [167] |
| Finnie-Ansley_2022 | The robots are coming: Exploring the implications of OpenAI codex on introductory programming | scopus | [104] |
| Kuramitsu_2024 | Training AI Model that Suggests Python Code from Student Requests in Natural Language | scopus | [195] |
| Shin_2024 | UNDERSTANDING OPTIMAL INTERACTIONS BETWEEN STUDENTS AND A CHATBOT DURING A PROGRAMMING TASK | scopus | [319] |
| Vishnu_2025 | Unveiling the Role of GPT-4 in Solving LeetCode Programming Problems | scopus | [351] |
| Brender_2024 | WhoÕs Helping Who? When Students Use ChatGPT toÊEngage inÊPractice Lab Sessions | scopus | [50] |
| Sun_2024 | Would ChatGPT-facilitated programming mode impact college studentsÕ programming behaviors, performances, and perceptions? An empirical study | scopus | [328] |
| Lohr_2025 | You're (Not) My Type-ÊCan LLMs Generate Feedback of Specific Types for Introductory Programming Tasks? | scopus | [210] |